\documentclass[twocolumn]{aastex61}

\usepackage{graphicx}
\usepackage{float}
\usepackage[caption=false]{subfig}
\usepackage{enumitem}

\newcommand{\degree}{^\circ}
\newcommand{\cloudUnit}{m$^2$/particle}

\received{March 22, 2018}
\revised{May 21, 2016}
\accepted{May 21, 2018}
\submitjournal{AJ}

\shorttitle{JWST Clear \& Cloudy Forecasts}
\shortauthors{Schlawin et al.}

\begin{document}

\title{Clear and Cloudy Exoplanet Forecasts for JWST: Maps, Retrieved Composition and Constraints on Formation with MIRI and NIRCam}

\correspondingauthor{Everett Schlawin}
\email{eas342 AT EMAIL Dot Arizona .edu}

\author{Everett Schlawin}
\affiliation{Steward Observatory \\
933 North Cherry Avenue \\
Tucson, AZ 85721, USA}

\author{Thomas P. Greene}
\affil{NASA Ames Research Center, Space Science and Astrobiology Division, Moffett Field, CA 94035, USA}

\author{Michael Line}
\affil{School of Earth and Space Exploration, Arizona State University, Tempe, AZ 85281, USA}

\author{Jonathan J. Fortney}
\affil{Department of Astronomy \& Astrophysics, University of California, Santa Cruz, CA, USA}

\author{Marcia Rieke}
\affiliation{Steward Observatory \\
933 North Cherry Avenue \\
Tucson, AZ 85721, USA}



\begin{abstract}

The James Webb Space Telescope (JWST) will measure exoplanet transmission and eclipse spectroscopy at un-precedented precisions to better understand planet structure, dynamics, chemistry and formation.
These are essential tools on the march \replaced{towards biosignatures}{toward biosignature} searches on potentially habitable planets.
We explore a range of exoplanet atmospheric conditions and forecast the expected results with JWST.
We take realistic CHIMERA models that match existing Spitzer and HST results and simulate the spectra achievable with the JWST MIRI + NIRCam Guaranteed Time Observations (GTO) survey, which includes observations of HD 189733 b, WASP-80 b, HAT-P-19 b, WASP-107~b, GJ 436 b and HAT-P-26 b.
We then retrieve atmospheric parameters from these spectra to estimate the precision to which the planets' atmospheric compositions can be measured.
We find that emission spectra have well-constrained unimodal solutions but transmission spectra near 10$\times$ solar abundance and solar C/O ratios can suffer from bimodal solutions.
Broad wavelength coverage as well as higher precision data can resolve bimodal solutions and provide dramatically better atmospheric parameter constraints.
We find that metallicities can be measured to within 20\% to 170\%, which approaches the precisions on Solar System planets, and
C/O ratios can be constrained to $\sim$10\% to 60\%, assuming that observers can leverage short wavelength data to select the correct solution from the bimodal posteriors.
These compositional precisions are sufficient to validate or refute predictions from disk formation models on final atmospheric abundances as long as their history is not erased by planet evolution processes.
We also show the extent to which eclipse mapping with JWST is possible on our brightest system HD~189733~b.

\end{abstract}




\section{Introduction} \label{sec:intro}

Transmission and emission spectroscopy reveal the compositions of an exoplanet atmosphere, beginning with the first atmospheric detection in HD 209458 b \citep{charbNa} and secondary eclipses of TrES-1 b and HD 209458 b \citep{charbonneau2005thermalEmission,deming2005secEclipse}.
After these initial discoveries, numerous ground-based and space-based observations of transiting exoplanets have revealed Na \citep[e.g.][]{huitson2012sodiumHD189}, K \citep[e.g.][]{sing11}, H$_2$O \citep[e.g.][]{deming13}, CO \citep[e.g.][]{snellen2008Na209} and recently evidence for TiO \citep[e.g.][]{evans2016wasp121H2OTiO,sedaghati2017tio,nugroho2017tioWasp33} in hot Jupiter atmospheres.
Beyond detection, the strength of spectral features can be used to measure the atmospheric abundances of individual atoms to build a chemical inventory of a planet's atmosphere.
Spectra provide insights on an atmosphere's temperature profile, vertical mixing and bulk composition.
By comparing the equilibrium expectations under the inferred temperature profile to the observed abundances, it is possible to infer disequilibrium chemistry and vertical mixing in exoplanet atmospheres, as has been done in the solar system \replaced{\citep{crossfield2015review}}{\citep[e.g.][]{prinn1977coConvection}}.
Furthermore, these spectra and inferred compositions can be used to compare different planets in so-called comparative exoplanetology.
This is where exoplanet science can expand our general understanding of planet formation and evolution, even if the measurements are far less comprehensive than for the solar system planets.

The Hubble Space Telescope (HST) has enabled high precision spectroscopy with the Wide Field Camera 3 (WFC3) grism mode from 1.1~$\mu$m to 1.7~$\mu$m \citep[e.g.][]{berta2012flat_gj1214}.
This bandpass encompasses a broad H$_2$O feature that is visible both in transmission and secondary eclipse spectra of exoplanets \citep[e.g.][]{deming13,line2016hd209wfc3,kreidberg2014wasp43}.
Atmospheric abundance retrievals are applied to these WFC3 measurements to provide the \replaced{oxygen abundance}{H$_2$O abundance} in planet atmospheres, a proxy for atmospheric metallicity if one assumes the elements heavier than Helium have constant number fractions relative to each other at near-solar values.
These studies have been used to constrain the planet-mass versus metallicity relation \citep[e.g.][]{kreidberg2014wasp43,line2016hd209wfc3,wakeford2017hatp26} as well as to study what factors contribute to the thickness of clouds in hot Jupiter atmospheres \citep{sing2016continuum}.

JWST's large collecting area \replaced{(6.5 m versus HST's 2.4 m)}{(25 m$^2$ versus HST's 4.5 m$^2$)} and longer wavelength capabilities will allow studies of the atmospheres of smaller and cooler transiting planets than what is achievable with HST.
Although most measurements of exoplanet atmospheres are of hot Jupiters since they are easier to study, {\it Kepler} has demonstrated that smaller planets $\lesssim 4 R_\oplus$ are far more common and that hot Jupiters are relatively rare beasts \citep{howard2012occurrenceKepler,petigura2017ckIVmetalRich}.
Atmospheric studies with JWST will therefore provide a broader understanding of smaller, more common, planets.
Pushing to cooler atmospheres will enable studies of CH$_4$ and NH$_3$ chemistry as well as approach habitable zone conditions such as in the TRAPPIST-1 system \citep{gillon2017trappist-1sevenp}.

JWST will dramatically expand the wavelengths available to measure carbon-bearing molecules like CO, CO$_2$ and CH$_4$ and nitrogen-bearing molecules like NH$_3$ \citep{greene2016jwst_trans}.
While HST already detected H$_2$O features, JWST will also perform detailed characterization of water vapor at higher precision.
The NIRISS (0.6-2.8$\mu$m), NIRCam (2.4-5$\mu$m), NIRSpec (0.6-5.2$\mu$m), MIRI Low Resolution Spectrometer (LRS) (5-12$\mu$m) instruments all provide time series modes for transmission and emission spectroscopy\footnote{See \url{https://jwst-docs.stsci.edu/display/JPP/Overview+of+Time-Series+Observation+\%28TSO\%29+Modes}} \citep[e.g.][]{beichman2014pasp} that will open up the wavelength space and precision available for exoplanet characterization.
Additionally, a proposed new mode for NIRCam makes use of a Dispersed Hartmann Sensor mode to add short wavelength coverage (1-2$\mu$m) to the NIRCam instrument simultaneously with long wavelength grism observations (2.4$\mu$m to 5.0$\mu$m) \citep{schlawin2017dhs,greene2016slitlessGrisms}.
JWST's large collecting area and uninterrupted observations of transiting systems will enable higher precision measurements than are possible with HST, which has to contend with an Earth occultation in each 96 minute orbital period.\footnote{\url{http://www.stsci.edu/hst/proposing/documents/primer/Ch\_64.html}}

Observations with these JWST modes will allow, for the first time, nearly full accounting of the carbon and oxygen content in exoplanet atmospheres.
The C/O ratio (ie. the relative abundance of carbon and oxygen) is an important diagnostics of formation that can indicate location relative to \replaced{disks}{disks'} ice lines \citep{oberg11,madhusudhan12,espinoza2017metalEnrichmentCtoO}.
One of the more recent findings connecting planet formation to the present-day derived atmospheric abundances highlights the importance of planetesimal accretion after initial formation.
These accreting planetesimals tend to lower the C/O ratios for most planets after their initial formation \citep{mordasini2016planetFormationSpec}.

In order to better perform comparative planetology and test predictions of planet formation, it is important to have a quantitative estimate of how well JWST will measure atmospheric parameters.
There are many different theoretical atmospheric models used in the literature including \added{NEMESIS \citep{irwin2008nemesis},} CHIMERA \citep{Line2013,line2014CtOsecE}, PETIT \citep{molliere2017jwst}, BART \citep{cubillos2016bart} or Exo-Transmit \citep{kempton2017exotransmit} that can simulate a transmission or emission spectrum of a planet.
The JWST signal to noise can be calculated for a set of observing parameters taking into account the orbital and stellar parameters such as in \citet{greene2016jwst_trans} and \citet{batalha2017pandexo}.
The high resolution forward model is then binned to the wavelengths of the observations and random noise is added to these spectra to simulate real JWST observations.
Finally, to measure the precision of atmospheric abundances and cloud properties, one can perform an atmospheric retrieval of the parameters using the simulated noisy spectrum and a Monte Carlo parameter estimation algorithm.

\citet{greene2016jwst_trans} perform these steps from a forward model to retrieved abundances for example cases: a hot Jupiter (HD 209458 b), warm Neptune (GJ 436 b), a warm sub-Neptune (GJ 1214 b) and cool super-Earth (K2-3 b).
These illustrative cases serve as guides to the expected precisions of a variety of planets with JWST.
In this paper, we extend the number of planets to demonstrate the capabilities of a Cycle 1 survey program and use more realistic models as a starting point.
We tune the models to match existing HST and Spitzer measurements so that our models are more realistic than the archetypes in \citet{greene2016jwst_trans}.
We show the expected results for the NIRCam and MIRI Guaranteed Time Observations (GTO) program to be carried out in JWST's Cycle 1.

Our calculations include several updates from \citet{greene2016jwst_trans}.
The instrument throughput and noise parameters for the telescope and instruments have been updated based on new models and measurements during cryogenic vacuum testing.
The CHIMERA models have been improved to include correlated-K abundances which speeds up opacity computations, equilibrium chemistry and a more flexible prescription for cloud opacity.
The assumption of equilibrium chemistry reduces the number of free parameters: instead of fitting each relevant molecule as a separate free parameter, the abundances and relative chemical ratios (C/O) are adjusted and then run through the CEA chemical equilibrium code\footnote{\url{http://www.grc.nasa.gov/WWW/CEAWeb/ceaHome.htm}}\citep{gordon1996cea} to predict CH$_4$, CO, CO$_2$, NH$_3$ and H$_2$O abundances.
This solves the problem of double-peaked priors of the C/O ratio and single-peaked or sloped priors in the metallicity that occur when assuming flat priors on individual molecules \citep{line2014CtOsecE}.
The newer code allows for cloud opacity at all pressure levels with a variable optical thickness and two parameters to describe the wavelength dependence: a short wavelength Raleigh-scattering portion of the spectrum and a turnoff to where it \replaced{because}{becomes} constant with wavelength.

\added{
Several previous studies have already explored information content in exoplanet spectra for retrievals \citep{batalha2017infoContent,howe2017informationJWST}, biases in atmospheric retrievals due to temperature profile assumptions \citep{rocchetto2016biasesJWST}  and systematic errors due to wavelength-stitching and star spots \citep{barstow2015jwstSystematics}.
In addition to these retrieval studies, several works have considered new effects in forward models like self-consistent temperature profiles (including clouds) for prime JWST targets \citep{molliere2017jwst} and the detectability of sulfur and phosphorous-bearing molecules \citep{wang2017h2sAndPH3}.
Our work complements these studies by calculating spectra using as-built JWST instrument and telescope parameters and JWST Astronomer Proposal Tool observing parameters of two accepted JWST programs.
We also tune the input models to match literature spectra from Spitzer and HST.
Here, we simulate forward models and perform atmospheric retrievals on the specific systems within the GTO NIRCam + MIRI surveys (programs \#1185 and \#1177).
}

We discuss the GTO NIRCam + MIRI  survey of warm to hot transiting planets in Section \ref{sec:survey} including how we set parameters for the planet models and generate simulated spectra.
We discuss the retrieval results in Section \ref{sec:results} and discuss their utility for comparative planetology and tests of formation models. 
In Section \ref{sec:EclipseMap}, we assess the viability of performing eclipse mapping of HD 189733 b.
Finally, we conclude in Section \ref{sec:Conclusion}.

\begin{deluxetable*}{ccccccccccccc}
\tabletypesize{\scriptsize}
\tablecaption{Model parameters and observing parameters of the planets included in the joint NIRCam + MIRI GTO Survey}
\tablehead{\colhead{Name} & \colhead{Mass} & \colhead{Radius} & \colhead{T$_{pl}$} & \colhead{T$_*$} & \colhead{Obs} & \colhead{NIRCam\tablenotemark{a}} & \colhead{MIRI} & \colhead{M/H} & \colhead{P(Q$_C$)} & \colhead{$\kappa_G$} & \colhead{C/O} & \colhead{$K_S$ Magnitude}\\
 & M$_\oplus$ & R$_\oplus$ & K & K &  & \# visits & \# visits & $\times$ solar & bar & \cloudUnit &  & (Vega) \\
}
\startdata
HD 189733 b & 360 & 12.5 & 1150 & 5040 & E & 2 & 0 & 3.5 & 0.001 &  & 0.55 & 5.54 \\
WASP-80 b & 180 & 10.7 & 900 & 4150 & E+T & 2 & 1 & 7.6 & 0.001 & $1 \times 10^{-29}$ & 0.55 & 8.35 \\
HAT-P-19 b & 93 & 12.7 & 1010 & 4990 & E & 2 & 1 & 15.5 & 0.001 &  & 0.55 & 10.55 \\
WASP-107 b & 38 & 10.5 & 770 & 4430 & T & 2 & 1 & 10 & 0.001 & $5 \times 10^{-29}$ & 0.03 & 8.64 \\
GJ 436 b & 22 & 4.2 & 700 & 3350 & E & 6 & 2 & 1000 & 0.32 &  & 0.55 & 6.07 \\
HAT-P-26 b & 19 & 6.2 & 990 & 5080 & T & 2 & 1 & 4.8 & 0.001 & $10^{-30}$ & 0.55 & 9.58\\
\enddata
\tablecomments{The GTO survey will include objects that have hot to warm temperatures (T$_{pl} \lesssim$ 1150 K) to study the emergence of methane (CH$_4$), which becomes energetically favored at these temperatures.
The adopted planet temperatures T$_{pl}$ are chosen to match secondary eclipse observations.
The ``Obs'' column lists whether the GTO program includes a transmission (T) and/or emission (E) spectrum of the target as well as the number of visits in MIRI and NIRCam.
The metallicity, M/H, is given as the multiple of solar abundance of heavy elements.
The carbon quench pressure, P(Q$_C$), indicates the pressure at which chemical reaction rates drop below vertical mixing rates.
Clouds are parameterized by a gray opacity times abundance, $\kappa_G$, due to large grains affecting transmission spectra only.
K magnitudes are from 2MASS \citep{skrutskie06}.
}\label{tab:gtoSources}
\tablenotetext{a}{At least two visits (one with the F444W and one with the F322W2 filter) are required to cover the 2.4$\mu$m to 5.0$\mu$m wavelength range.}
\end{deluxetable*}

\section{A Planet Survey from warm to hot planets}\label{sec:survey}

The combined NIRCam and MIRI Guaranteed Time Observations (GTOs) include a dedicated spectroscopic survey of transiting planets across a variety of sub-Neptune to Jupiter sized planets that live in warm ($T_{eq} \sim$650 K) to hot ($T_{eq} \sim$1200K) environs, where $T_{eq}$ is the zero-albedo equilibrium temperature.
The survey will extend the previous trends hinted at in \citet{stevenson2016cloudPresence} and \citet{crossfield2017neptuneTrends} while also adding valuable data on carbon-bearing molecules such as methane, carbon dioxide and carbon monoxide.

\citet{greene2016jwst_trans} modeled the transmission spectra and emission spectra of archetypical planets to assess the best JWST modes and targets to achieve high precision atmospheric abundances and cloud constraints using the CHIMERA model suite \citep{Line2013,line2014CtOsecE}.
An important finding from this initial result is that the synergy between different instruments is important to cover wavelength space and abundance degeneracies.
This is why the NIRCam GTO observations (2.4~$\mu$m to 5.0~$\mu$m) from JWST Program \#1185 will be combined with MIRI GTO observations (5.0~$\mu$m to 11~$\mu$m) from JWST Program \#1177 into a single data set for joint analysis.
For most of our targets, NIRISS data will also be obtained by a different GTO program, enabling a complete (0.6~$\mu$m to 11~$\mu$m) census on molecules and clouds in these atmospheres.
We discuss our model planet parameters for this GTO survey using previous observations as a guide.
Table \ref{tab:gtoSources} lists the properties of the planets in our survey as well as the adopted parameters for the atmospheric models.

\subsection{Observing Summary}
For the NIRCam grism time series observations, light is dispersed with a grism in the pupil wheel and the wavelengths are selected in the filter wheel with either a F322W2 (2.43 -- 4.01~$\mu$m) or F444W (3.88 -- 4.99~$\mu$m) wide band filter on separate visits \citep{greene2017jatisNIRCam}.
The NIRCam observations will also make use of simultaneous short wavelength photometry in the F210M band (2.0 -- 2.2~$\mu$m) that is spread over many pixels with either a +4 or +8 wave weak lens in the pupil wheel.
These short wavelength data can be used to monitor the centroid and help correct for systematics due to stellar activity.
For the MIRI LRS observations, the light is dispersed by a prism to cover the wavelength range from 5.0 to 11 $\mu$m \citep{kendrew2015LRSMIRI}.
Although MIRI LRS continues beyond 11~$\mu$m, we cut off the longest wavelength at 11~$\mu$m because the sensitivity drops significantly with wavelength.
For both the NIRCam and MIRI observations, there is no slit to minimize systematics from slit losses due to pointing or centering in a silt.
The tradeoff is that the position angle constraints must be designed to \replaced{ensure no}{minimize} contamination of the spectrum from nearby stars.

We assume the parameters of the submitted GTO programs \#1185 and \#1177 employing observations with the NIRCam F322W2 and F444W time series grism as well as the MIRI LRS.\footnote{For the detailed observing specifications and proposal files, see \url{https://jwst-docs.stsci.edu/display/JSP/JWST+GTO+Observation+Specifications}}
As listed in Table \ref{tab:gtoSources}, the combined programs include transit observations of 3 planets and eclipse observations of 4 planets, with WASP-80 b in common with both \replaced{lists}{observation types} to ensure one planet will be well characterized from both geometries.
All planets, except for HD 189733~b, will be visited for at least 3 separate transits/eclipses to cover 3 different wavelength regions: a) NIRCam F322W2 , b) NIRCam F444W and c) MIRI LRS.
This ensures complete wavelength coverage from 2.4 $\mu$m to 11$\mu$m for these sources.
HD 189733 b is too bright for the MIRI LRS mode so it will only be observed with the a) NIRCam F322W2 and b) NIRCam F444W modes.
Each planet will be visited for a single transit and/or eclipse in each wavelength mode except for GJ 436 b where extra signal to noise is desired.
GJ 436 b eclipses will be observed 3 times in each NIRCam filter and 2 times \replaced{in the}{with} MIRI for a total of 8 separate visits during eclipses.
The combined photon collection time for all observations listed in Table \ref{tab:gtoSources} is 137 hours, which is charged as 190 JWST hours including all overheads, representing a science efficiency of 72\%.

For some example cases discussed in Section \ref{sec:twoModes}, we also consider the effect of adding NIRISS observations with the Single Object Slitless (SOSS) instrument throughputs, as in \citet{greene2016jwst_trans} for the wavelengths from 1 to 2.4~$\mu$m.
We include only this subset of the maximum possible wavelength range because the sensitivity drops below $\sim$1~$\mu$m and because the spectral orders overlap above $\sim$2.4~$\mu$m.
The NIRISS data pipeline may efficiently remove the second order contamination with exposures taken with different filter combinations, but we do not consider the effect of additional wavelength coverage in this work.

\subsection{Emission and Transmission Spectra}\label{sec:emisTransSpec}
\begin{figure*}
\gridline{\fig{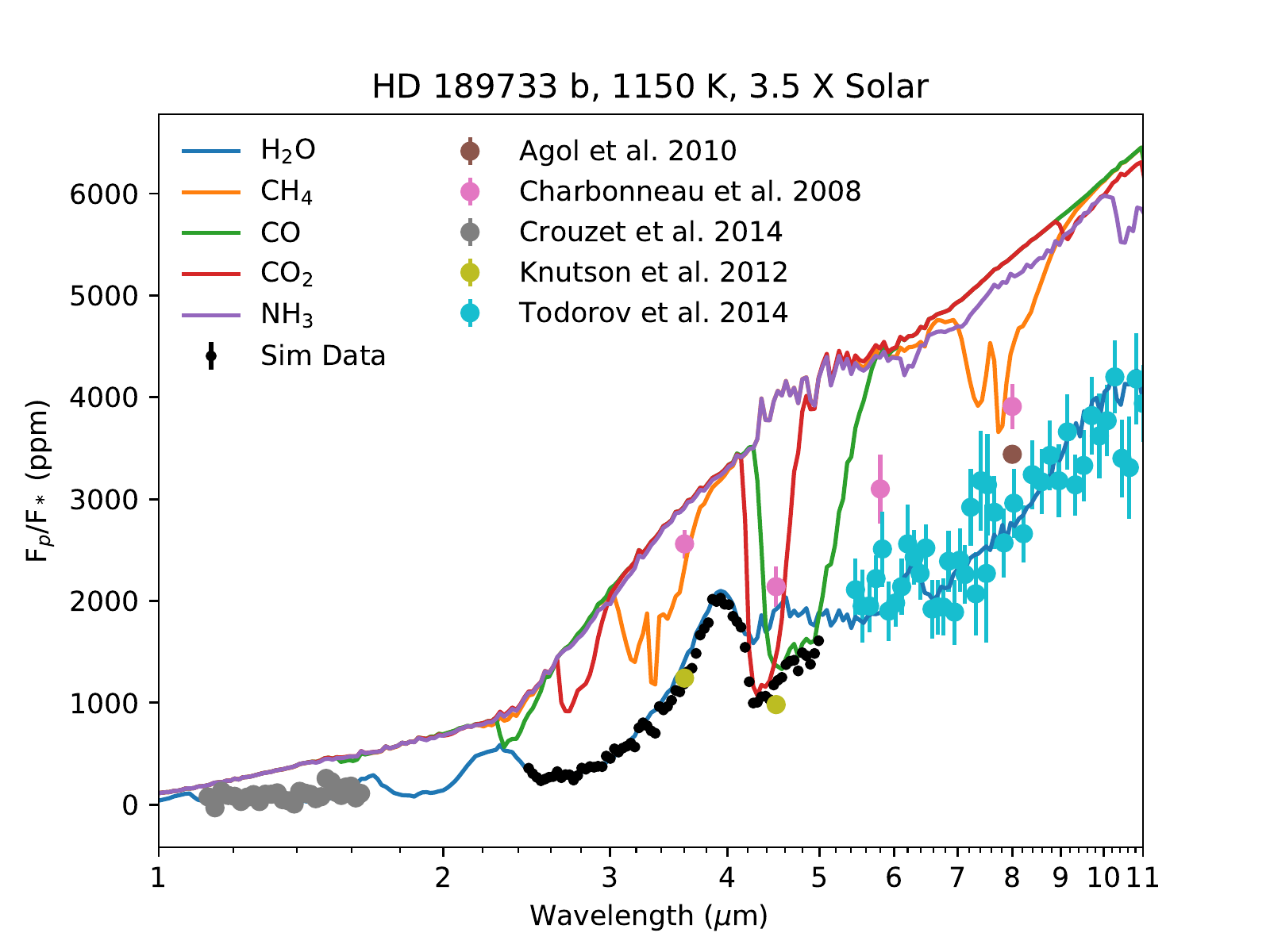}{0.5\textwidth}{HD 189733 b (360 M$_\oplus$) Emission}
          \fig{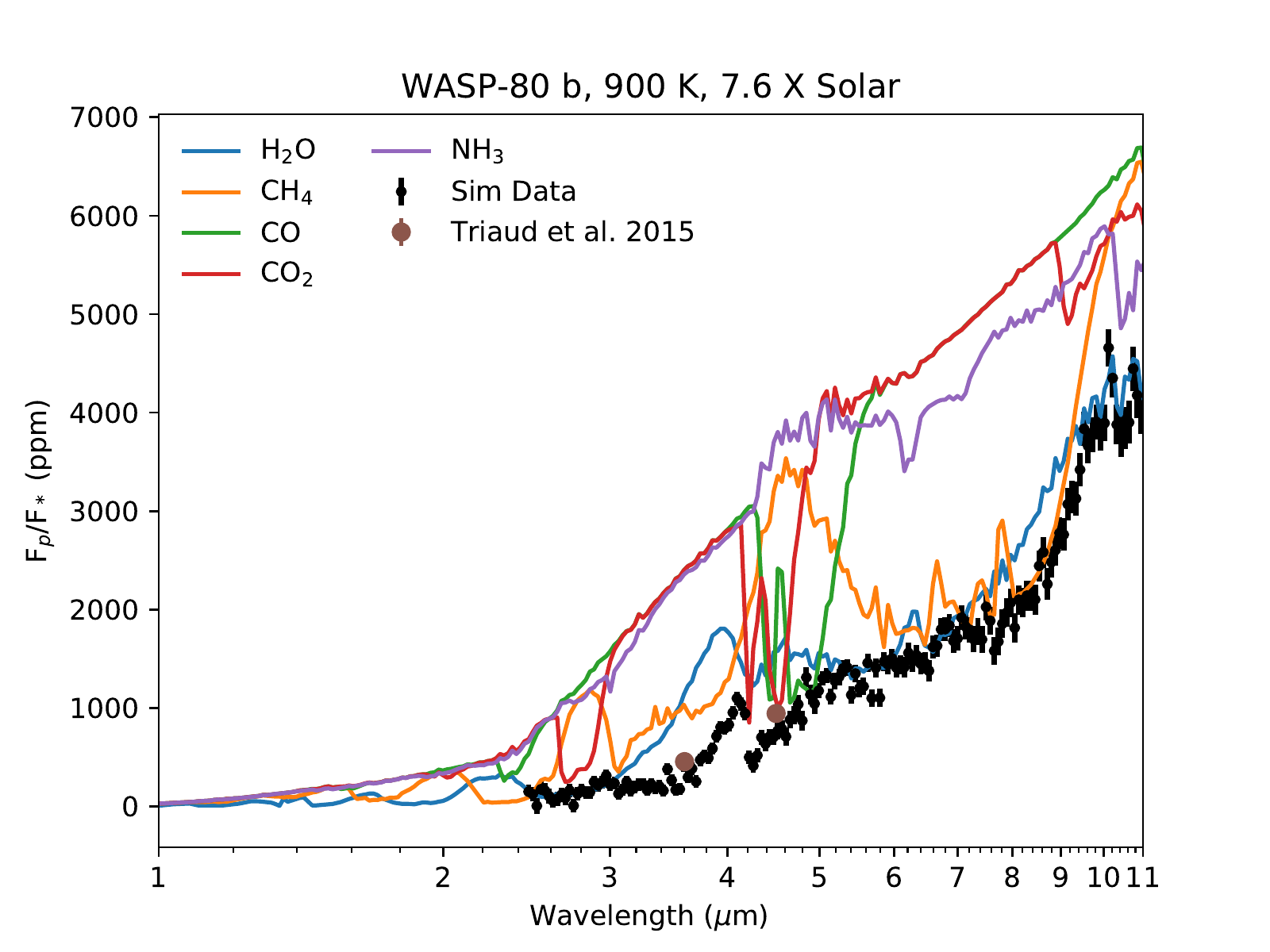}{0.5\textwidth}{WASP-80 b (180 M$_\oplus$) Emission}          }
\gridline{\fig{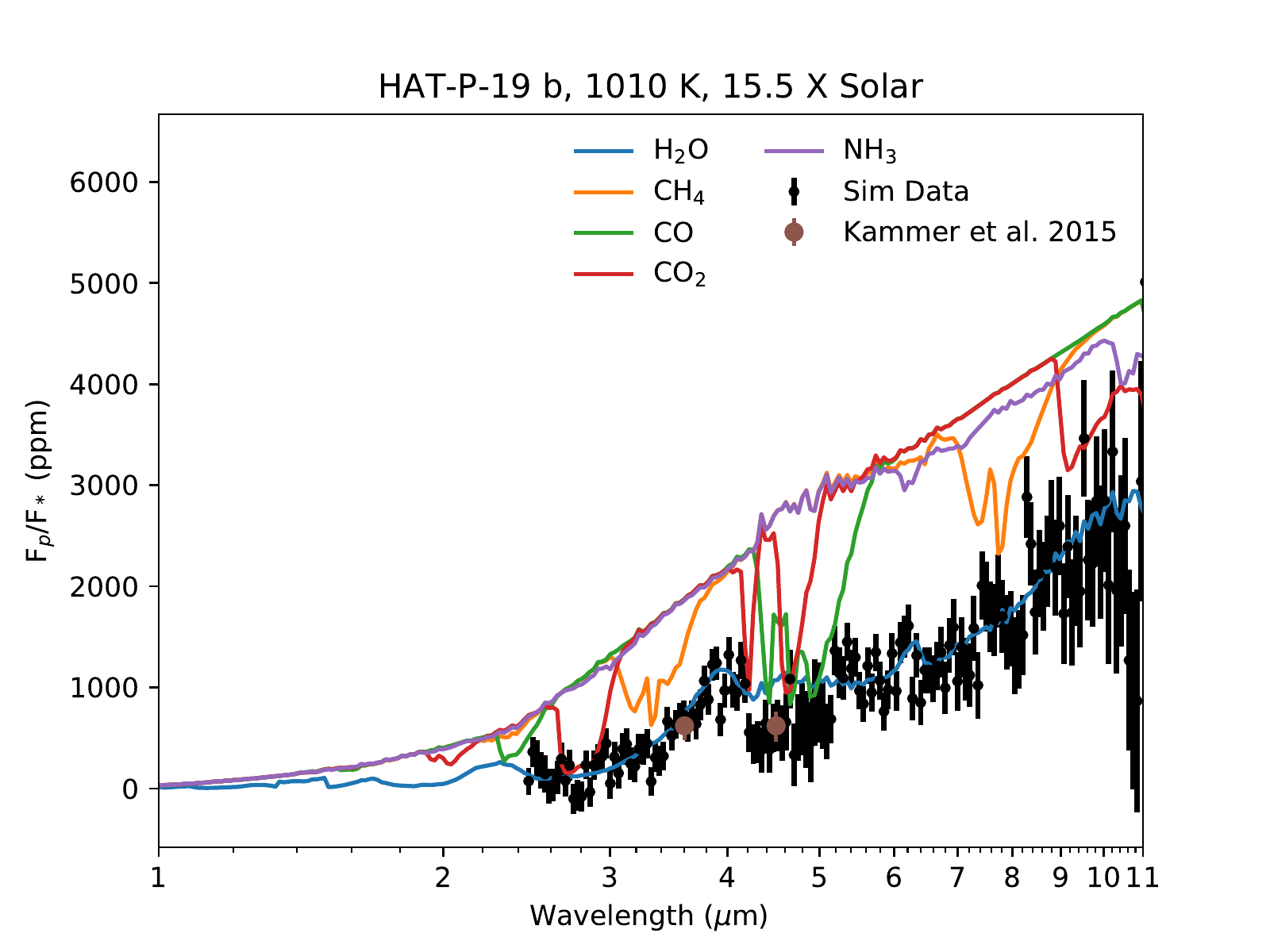}{0.5\textwidth}{HAT-P-19 b (93 M$_\oplus$) Emission}
	 \fig{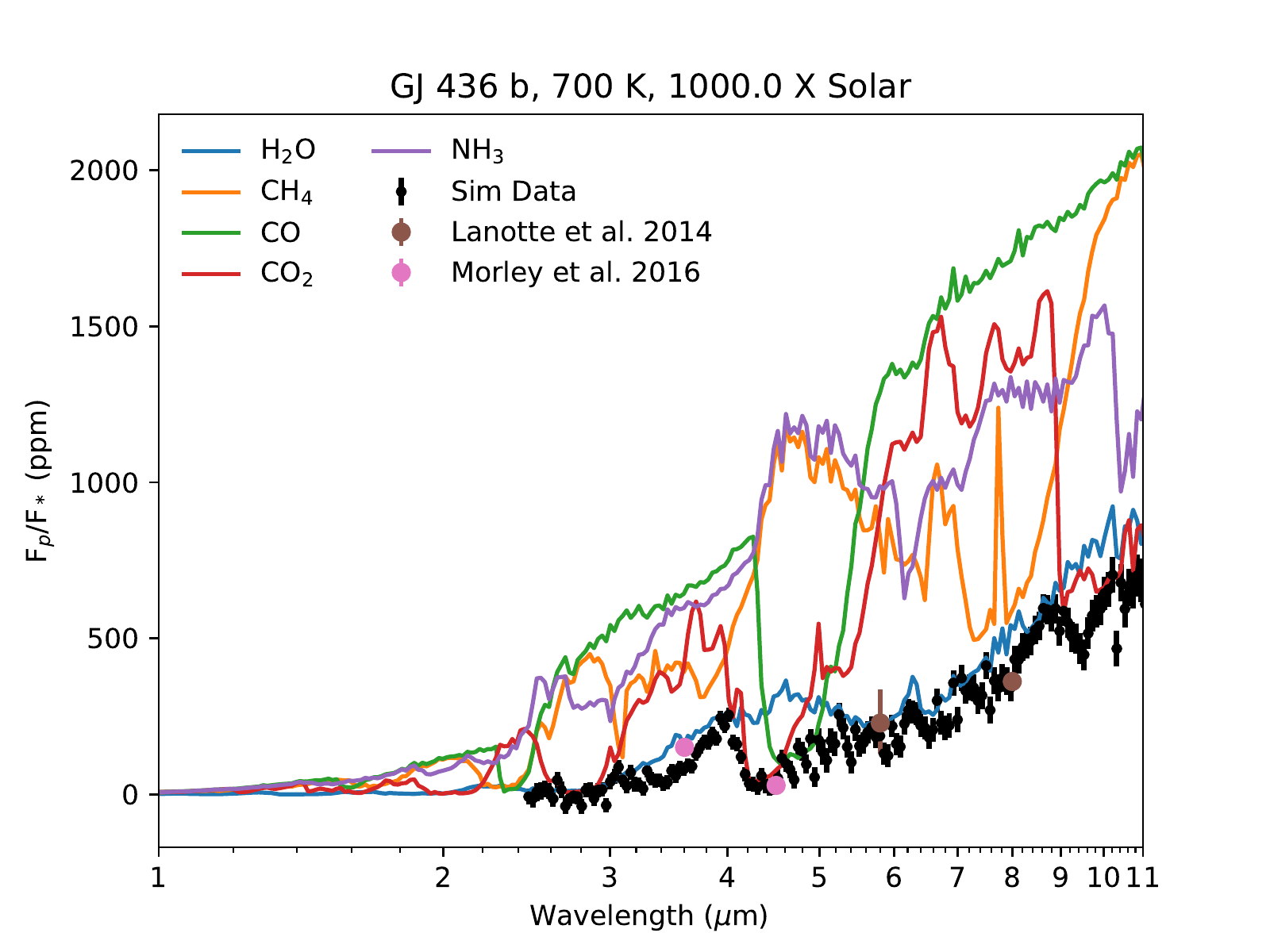}{0.5\textwidth}{GJ-436 b (22 M$_\oplus$) Emission}	}
\caption{We show our theoretical emission spectra for the GTO targets with simulated JWST data with black error bars.
The contributions from significant absorbers in the emission spectrum are shown as multi-colored lines.
These contributions are calculated by creating models with the same temperature profile and abundances as the full model but where all gases are made artificially transparent, except for collisionally-induced absorption opacity and the one being isolated.
Over-plotted on the spectra are previous measurements from \citet{agol2010hd189},
\citet{charbonneau2008hd189emission}, \citet{crouzet2014water189}, \citet{knutson2012phase189},
\citet{todorov2014hd189IRS}, \citet{triaud2015wasp80Dayside}, \citet{kammer2015SE5giants},
\citet{lanotte2014gj436} and \citet{morley2017gj436bRetrieval}.
The models were adjusted to be consistent with existing measurements where possible, which required an unusually high metallicity and quenching carbon chemistry at 0.32 bar for GJ 436 b.}\label{fig:emisOpacities}
\end{figure*}

\begin{figure*}
\gridline{          \fig{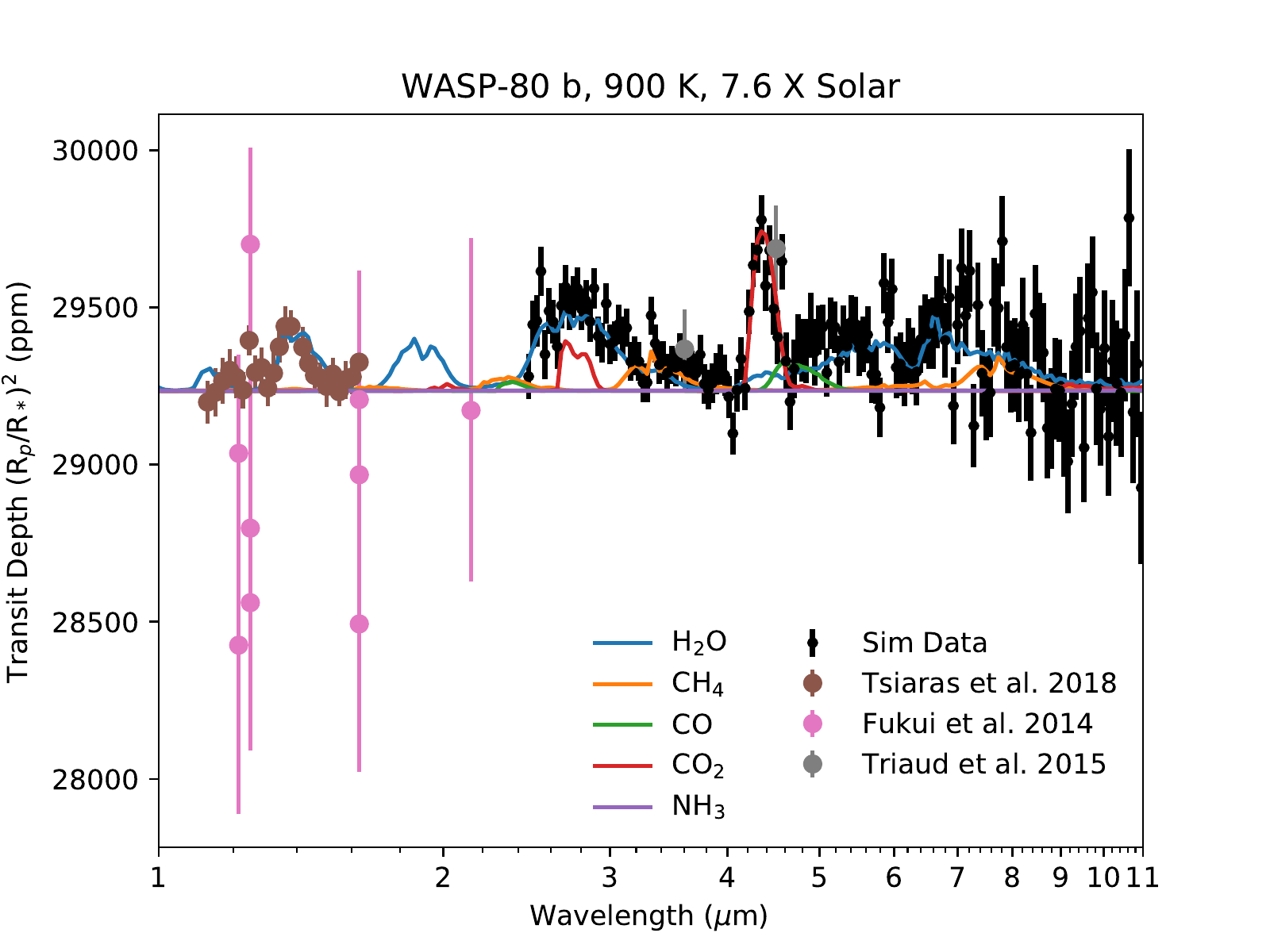}{0.5\textwidth}{WASP-80 b (180 M$_\oplus$) Transmission}          }
\gridline{\fig{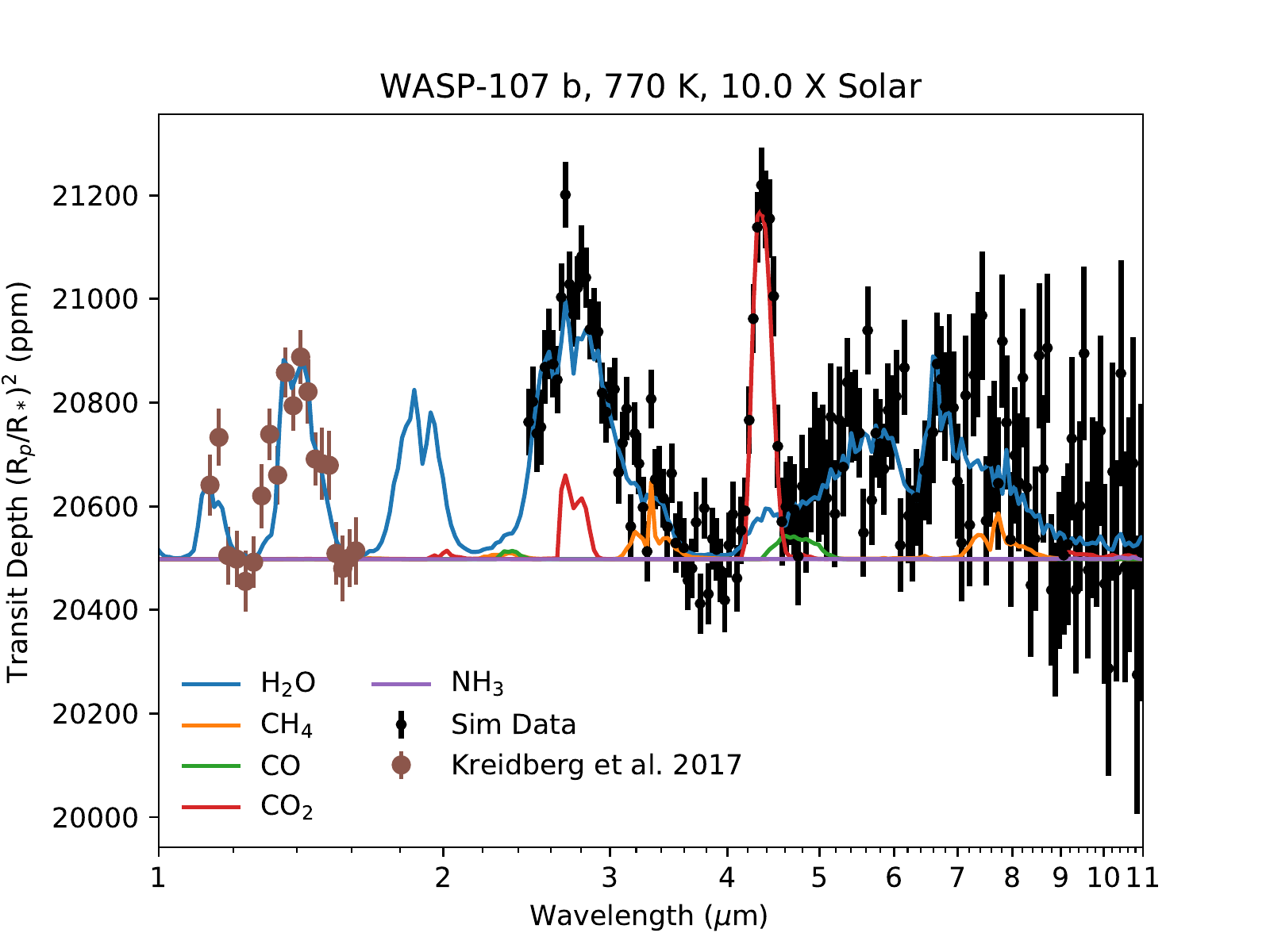}{0.5\textwidth}{WASP-107 b (38 M$_\oplus$) Transmission}
	 \fig{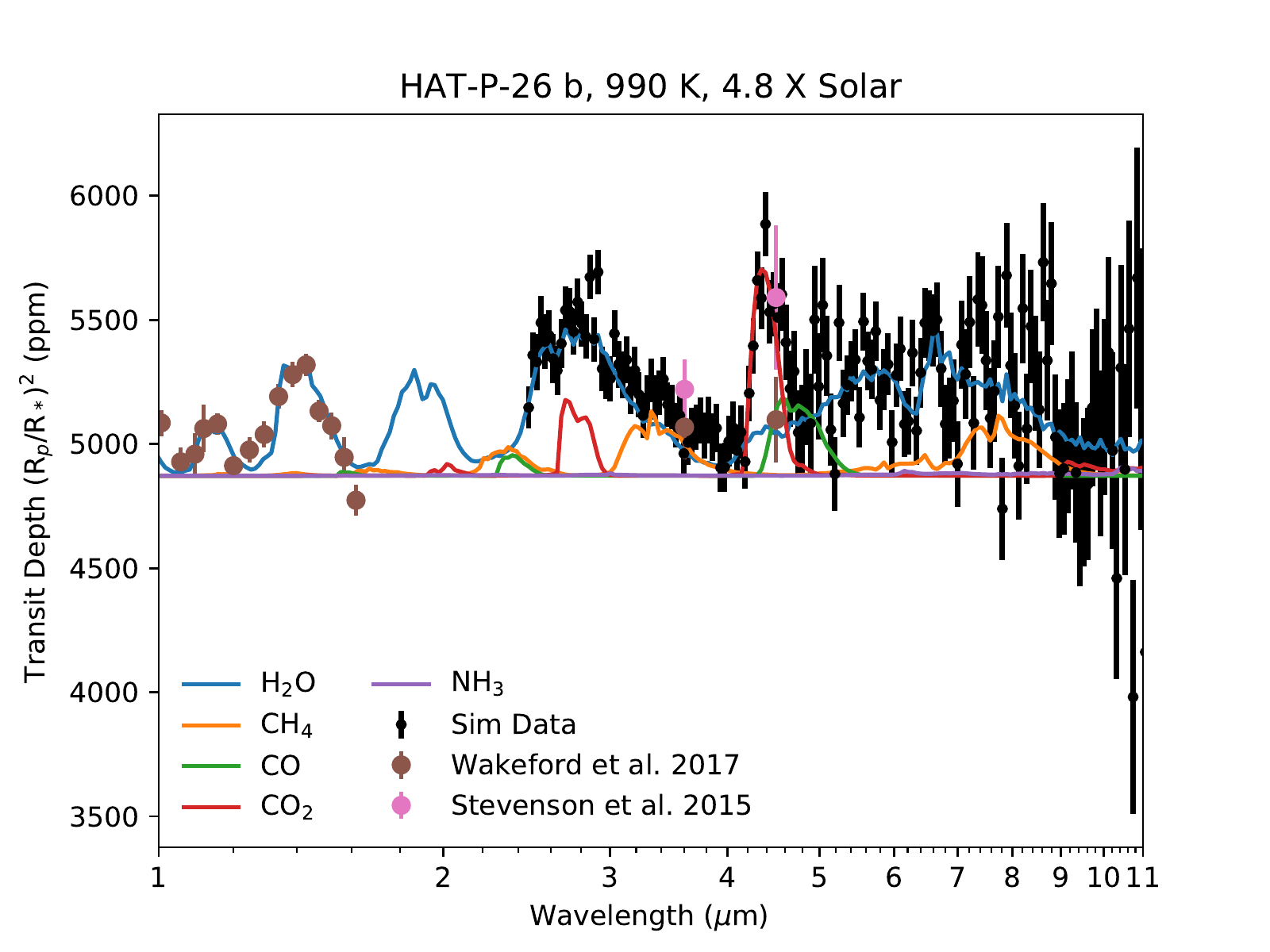}{0.5\textwidth}{HAT-P-26 b (19 M$_\oplus$) Transmission}	}
\caption{We show our theoretical transmission spectra for the GTO targets with simulated JWST data with black error bars.
The contributions from significant absorbers in the transmission spectrum are shown as multi-colored lines.
These contributions are calculated by creating models with the same temperature profile and abundances as the full model but where all gases are made artificially transparent, leaving just the opacity of clouds and the molecule being isolated.
Over-plotted on the spectra are previous measurements from \citet{tsiaras2018hstHotJups},
\citet{fukui2014wasp80}, \citet{triaud2015wasp80Dayside}, \citet{kreidberg2017wasp107},
\citet{wakeford2017hatp26}, and \citet{stevenson2016hatp26}.
The models were adjusted to be consistent with existing measurements where possible, which included a gray cloud opacity and a low C/O ratio for WASP-107 b to be consistent with \citet{kreidberg2017wasp107}.}\label{fig:transOpacities}
\end{figure*}

We collected high precision spectra of the planets in our GTO survey that were available in the literature and create initial models to be consistent with these data.
For HD 189733 b, we include secondary eclipse measurements from \citet{agol2010hd189}, \citet{charbonneau2008hd189emission},
\citet{knutson2012phase189} and \citet{crouzet2014water189}.
For the Spitzer IRS spectrum of HD 189733 b, we use the more recent re-analysis of the \citet{grillmair2008water} result performed by \citet{todorov2014hd189IRS}.
We adjust the CHIMERA model to match the later (post 2011) observations shown in Figure \ref{fig:emisOpacities} since the methods to correct for Spitzer systematics have improved over time \citep{ingalls2016spitzerRepeatability}.
For WASP-80 b, we use the secondary eclipse values from \citet{triaud2015wasp80Dayside}.
For WASP-80 b's transmission spectrum, we use ground-based measurements from \citet{fukui2014wasp80}, Spitzer measurements \citet{triaud2015wasp80Dayside} and \added{the} HST spectrum from \citet{tsiaras2018hstHotJups}.
For HAT-P-19 b, we use the secondary eclipse spectrum from \citet{kammer2015SE5giants}.

For GJ 436 b, we use secondary eclipse measurements from \citet{lanotte2014gj436} and \citet{morley2017gj436bRetrieval}.
As found in \citet{morley2017gj436bRetrieval}, the high 3.6~$\mu$m to 4.5~$\mu$m flux ratio requires that methane be suppressed in the atmosphere.
We force the metallicity of GJ 436 b to be high \replaced{(1000$\times$ solar)}{\citep[1000$\times$ solar][]{morley2017gj436bRetrieval,moses2013compositionalDiversity}} and also insert a carbon quench pressure of 0.32 bar, which ensures that the low CH$_4$ mixing ratio from deeper altitudes persists to higher elevations.
For WASP-107 b, we use the transmission spectrum from \citet{kreidberg2017wasp107}, which also has suppressed CH$_4$ features.
To be consistent with the \citet{kreidberg2017wasp107} retrievals, we assume a gray cloud opacity to reduce the depth of the H$_2$O features as well as a low C/O ratio to reduce CH$_4$ opacity.
Though the methane is undetectable in the HST bands with this model, it may reveal a small feature at 3.3~$\mu$m detectable in the F322W2 NIRCam grism mode shown in Figure \ref{fig:transOpacities}.
For HAT-P-26 b, we use the transmission spectrum from \citet{stevenson2016hatp26} and \citet{wakeford2017hatp26}.

\subsection{Planet Metallicity and Abundances}\label{sec:abundMetal}

We assume that the planet metallicity follows the mass-metallicity trend in \citet{kreidberg2014wasp43} for solar system planets and exoplanets or use literature values where necessary.
This trend is approximated as
\begin{equation}\label{eq:MassMetal}
[M/H] = -1.1 \log{\left(M / M_{jup}\right)} + 0.6,
\end{equation}
where $[M/H]$ is the log base 10 of the metallicity divided by solar abundance and $M$ is the planet mass.
This trend, though based on a small number of objects, also correctly predicts the metallicity of WASP-12 b \citep{kreidberg2015wasp12} but not GJ 436 b, WASP-107 b and HAT-P-26 b, as discussed below.

Population synthesis models show that there can be a diversity of planet envelope metallicities during planet formation \citep[e.g.][]{fortney2013lowMassP}, so Equation \ref{eq:MassMetal} should be thought of as \replaced{an average}{a guide} rather than \added{an} exact prescription.
For GJ 436 b, a high metallicity and a chemical quench point are needed to suppress CH$_4$ absorption to match the Spitzer secondary eclipse measurements \citep{morley2017gj436bRetrieval}.
For WASP-107 b, internal structure models limit the maximum metallicity and we adopt 10$\times$ solar.
The Neptune-mass planet HAT-P-26 b shows a metallicity below the \citet{kreidberg2014wasp43} trend, possibly because it formed close to its host star or late in the disk lifetime \citep{wakeford2017hatp26}.
We adopt the \citet{wakeford2017hatp26}-derived metallicity of 4.8$\times$ solar.
We assume solar C/O ratios (C/O = 0.55) for all planets in the GTO survey except for WASP-107 b, which requires a low C/O of $\sim$ 0.03 to explain the small CH$_4$ features in its transmission spectrum.

\subsection{Disequilibrium Chemistry}
\replaced{As vertical mixing rates exceed the chemical reaction timescales, the relative abundances of molecules can begin to deviate from equilibrium expectations.}{When the chemical reaction timescales (to reestablish chemical equilibrium) exceed the vertical mixing timescales, abundances can deviate from their equilibrium values. 
Instead, the abundances will resemble the equilibrium from lower altitudes (higher pressures) where chemical reaction timescales are shorter \citep[e.g.][]{visscher11}.}
We parameterize this with a chemical quench level P(Q), above which the atmospheric abundances are locked in the same ratios as the quench point.
For Nitrogen reactions, we assume equilibrium and set this point at P(Q$_N$) = 10$^{-4}$ bar, essentially above the photosphere where quenching will not affect the spectrum.
For Carbon reactions, we assume equilibrium (P(Q$_C$) = 10$^{-4}$ bar) for all of the planets except for GJ 436 b, where disequilibrium is needed.
Chemical quenching is required to explain the unusually high 3.6 $\mu$m Spitzer secondary eclipse which would otherwise be absorbed by CH$_4$ \citep{morley2017gj436bRetrieval}.
We set P(Q$_C$) at 0.32 bar to quench the carbon chemistry with a relatively high CO/CH$_4$ ratio to match these observations and reduce the CH$_4$ absorption.

\subsection{Clouds and Hazes}\label{sec:Clouds}
We assume that clouds in all transmission spectra can be parameterized by $\kappa_G$, which is the abundance-weighted cross section of large grain sizes in \cloudUnit.
These large grains are assumed to have radii $\gtrsim$10 $\mu$m so that their cross sections have no dependence on wavelength across JWST time series wavelengths.
In other words, the clouds produce a gray (constant with wavelength) opacity.
This parameterization can produce flat featureless spectra found on many exoplanets observed in transmission such as GJ 436 b \citep{knutson2014gj436}.
For HAT-P-26 b, we assume $\kappa_G = 10^{-30}$ \cloudUnit\ because  \citet{wakeford2017hatp26} find a clear atmosphere.
WASP-107 b's transmission spectrum has smaller atmospheric features than would be expected for this low surface gravity \replaced{38 $M_{jup}$}{$g$ = 3.4 m/s$^2$} planet \citep{kreidberg2017wasp107} so we include a cloud opacity of $5 \times 10^{-29}$~\cloudUnit\ to best-match the data.
Similarly, WASP-80 b requires a small level of cloud opacity at 1$\times 10^{-29}$~\cloudUnit\ to best match the HST spectrum from \citet{tsiaras2018hstHotJups}.
For the emission spectra, we do not include the effects of clouds since they are viewed nearly perpendicular to the line of sight and do not obtain the same optical depths as transmission looking at slant paths \citep{fortney2005condensates}.

Optical spectra show a Raleigh-like haze slope in the transmission spectra of many exoplanets \citep[e.g.][]{sing2016continuum}.
Therefore, our models have a Raleigh scattering slope and amplitude parameters.
Our survey focuses on the near and mid-infrared, so we assume the haze slope is negligible for these wavelengths in the forward models.

\subsection{Temperature Profile}\label{sec:TPprofile}
We assume that the temperature profile of these GTO targets is described by a 5 parameter analytic model from \citet{Line2013} and \citet{guillot2010radEquilibrium}.
In all cases, we assume a negligible internal heat flux of T$_{int}$ = 100 K.
For the infrared opacity $\kappa_{IR}$, we assume 3.2$\times 10^{-3}$ cm$^2$/g.
For the mean opacities of the two streams we assume $\gamma_1$ = $\gamma_2$ = 0.1.
We assume a re-radiation + albedo factor $\beta$ of 1.0 and equal partitioning of the streams so that $\alpha$ = 0.5.
The one exception is GJ 436, for which we adjust the profile to have $\gamma_1$=0.03 and $\gamma_2$=0.1 to best match the secondary eclipse spectrum from \citet{morley2017gj436bRetrieval}.

\added{
\subsection{Signal to Noise Estimates}
Once we adopted these abundance, quenching and temperature profile parameters and calculate the forward models for the planets in the program, we created simulated JWST spectra for each mode listed in Table \ref{tab:gtoSources}.
The simulated spectra were calculated for R=100 with the same code and methodology as in \citet{greene2016jwst_trans} with the following modifications:
1) We use a newer instrument throughput file for the NIRCam module A\footnote{\url{https://jwst-docs.stsci.edu/display/JTI/NIRCam+Filters}} which lowers the combined telescope + instrument + quantum efficiency throughput by 0 to 5\% depending on the wavelength.
2) We used a larger NIRCam aperture with 15 instead of 6 spatial pixels in radius, which increases the extracted flux from 90-98\% and decreases the noise impact of any imperfect centering. 
3) We increased the MIRI LRS background flux from 96 e$^-$/(s px) to 169 e$^-$/(s px) to be consistent with \citet{glasse2015sensitivity}, which reduces the signal the noise at the longest wavelengths  ($\gtrsim 9~\mu$m).
Overall, these changes should result in simulations that are closer to the expected JWST on-orbit performance

Studying the atmosphere of transiting exoplanets requires extreme precisions ($\lesssim 100$ppm photometric uncertainties on transit/eclipse depths), which means that systematic errors can raise the noise above the photon noise limit (determined by Poisson statistics of arriving photons and read noise).
These systematics errors include both astrophysical effects (such as star spots) and instrumental effects (such as charge trapping and thermal breathing) that can impact light curves \citep[e.g.][]{beichman2014pasp,barstow2015jwstSystematics,wakeford2016marginalizingSys,zhou2017chargeTrap}.
As in \citet{greene2016jwst_trans}, we include a noise floor that approximates these systematics that is added in quadrature to the photon and read noise contributions.
We assume a constant 30 ppm noise floor on the transit/eclipse depth for the NIRCam grism mode, which uses an HgCdTe detector that may have systematics resembling HST's WFC3 instrument.
We assume 50 ppm for MIRI, which has a SiAs detector that may have systematics resembling the {\it Spitzer} IRAC instrument.
These are the same noise floors that went into \citet{greene2016jwst_trans}.
}

\subsection{CHIMERA Retrieval Models}\label{sec:modelDescrip}

We retrieved model parameters for all simulated spectra in the GTO survey described in Section \ref{sec:survey} using CHIMERA with the \texttt{multinest} Bayesian inference tool \citep{feroz2009multinest}.
This tool has the advantage over traditional MCMC \citep[e.g.][]{foreman-mackey2013emcee} because it requires fewer likelihood evaluations in assessing posterior distributions, less sensitivity to the tuning of parameters and more straightforward convergence testing \citep{feroz2009multinest}.

The transmission retrievals used 9 free parameters: the 1) temperature (T), 2) metallicity (M/H), 3) carbon to oxygen ratio (C/O), 4) Raleigh haze amplitude (Amp$_{ray}$), 5) Raleigh haze slope (Slope$_{ray}$), 6) the large grain cloud opacity ($\kappa_{Cloud}$) discussed in Section \ref{sec:Clouds} 7) the 10-bar radius of the planet as a fraction of observed radius (x$_{R_p}$) 8) the first stream profile parameter (log($\gamma_1$)) and 9) Infrared opacity (log($\kappa_{IR}$)). Parameters 8) and 9) are used to set the temperature-pressure profile discussed in Section \ref{sec:TPprofile}.
The emission retrievals used 7 free parameters: the 1) first stream parameter (log($\gamma_1$)), 2) second stream parameter (log($\gamma_2$)), 3) infrared opacity (log($\kappa_{IR}$)), 4) two stream partitioning parameter ($\alpha$), 5) the equilibrium temperature (T), 6) the metallicity (M/H) and 7) the carbon to oxygen ratio (C/O).

We fixed the carbon and nitrogen quenching parameters at 10$^{-4}$ bar, which keeps everything in chemical equilibrium at the relevant pressures.
This means that our atmospheric retrievals will not converge on the exact solution for GJ 436 b since the forward model used to simulate the spectrum does include carbon quenching for GJ 436 b.

\begin{figure*}
\gridline{\fig{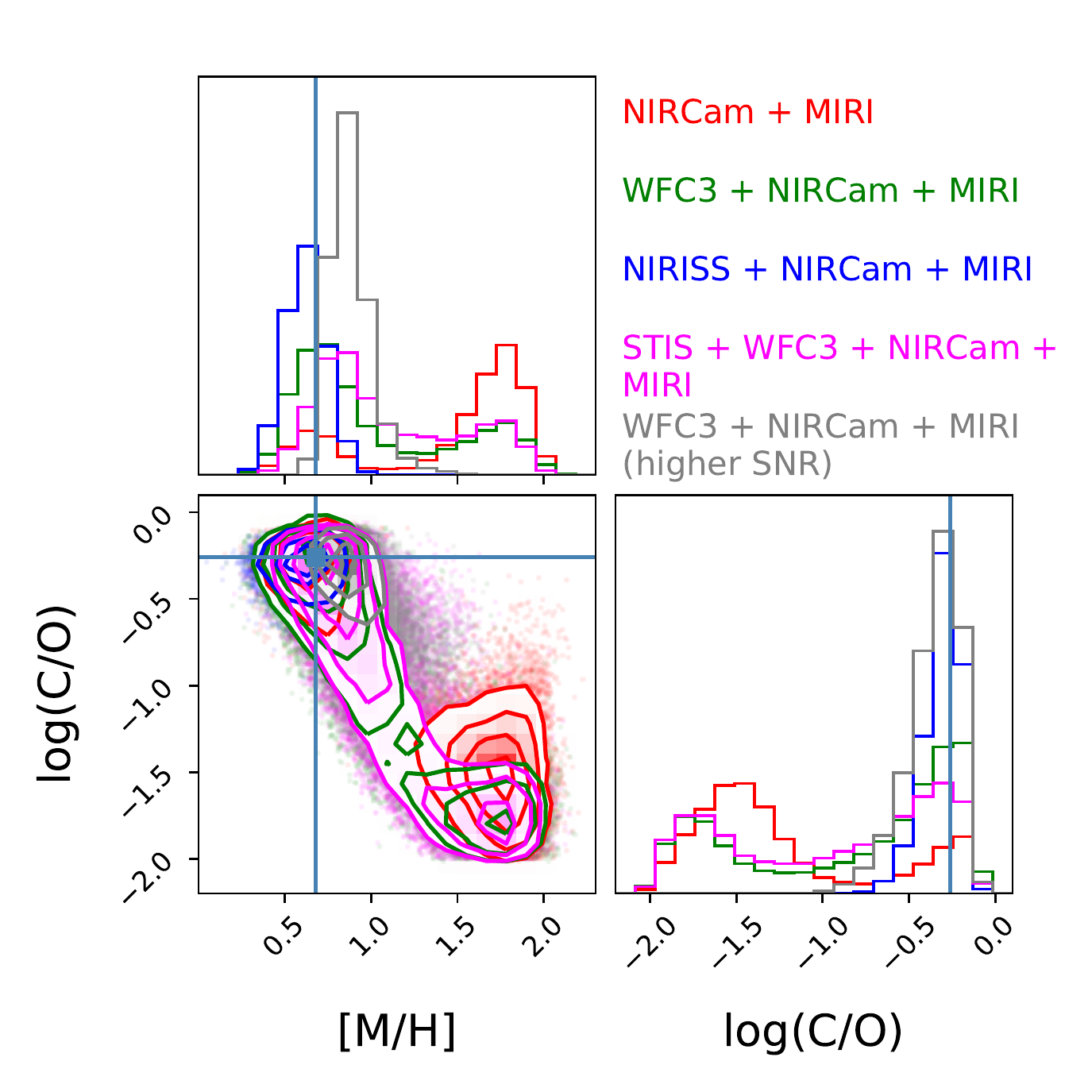}{0.32\textwidth}{HAT-P-26 b}
	    \fig{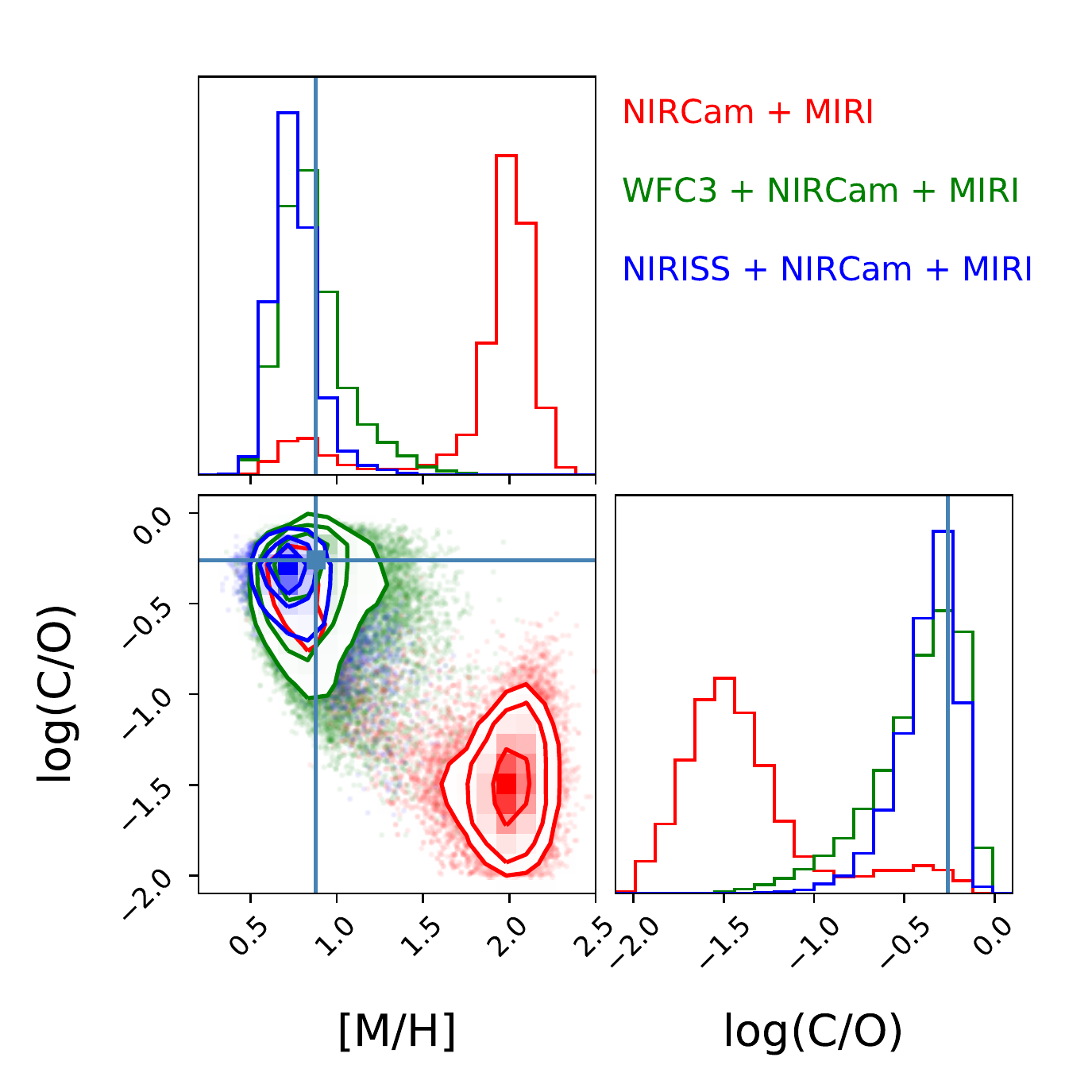}{0.32\textwidth}{WASP-80b}
   	    \fig{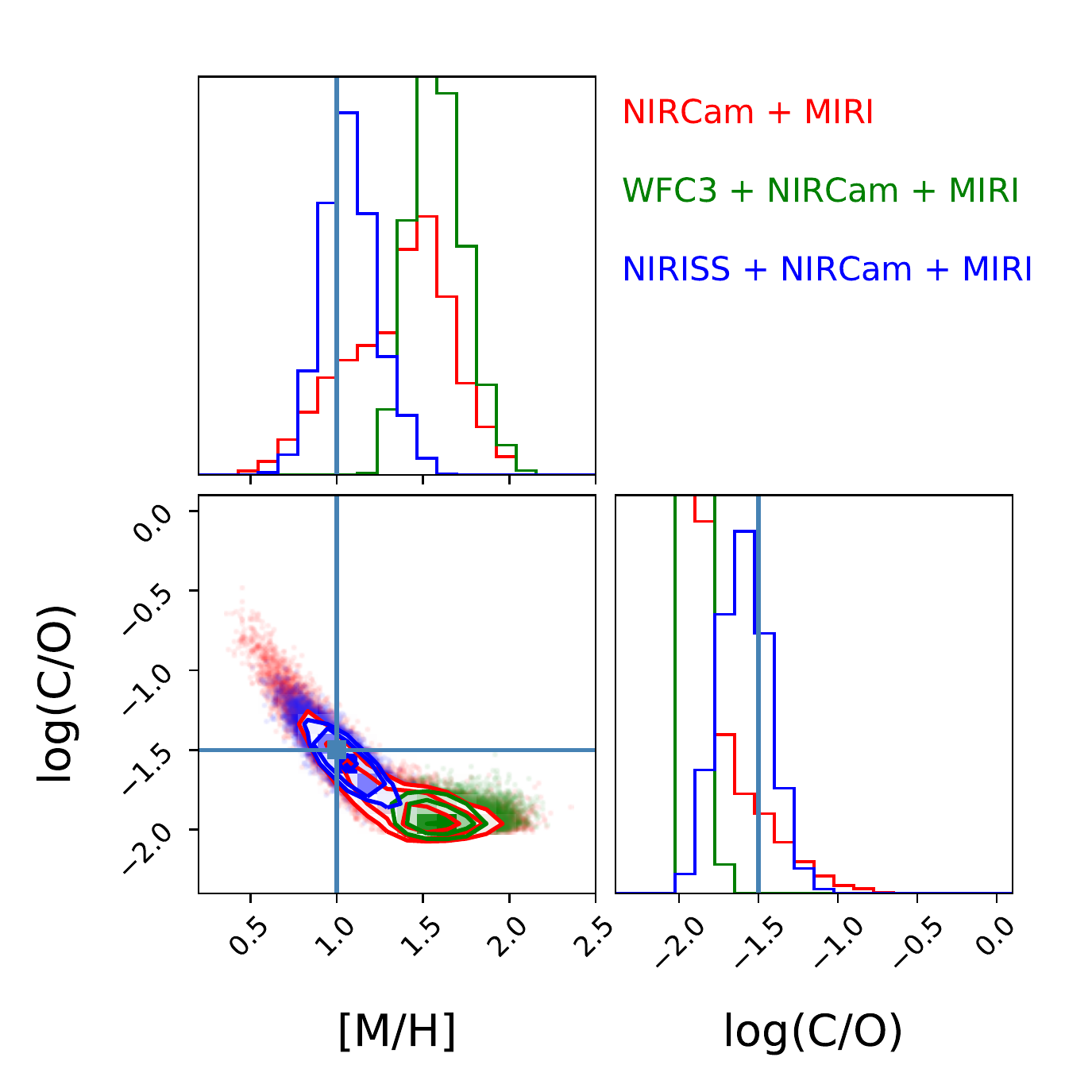}{0.32\textwidth}{WASP-107b}
}
\caption{Two mode solutions are retrieved when measuring a transmission spectrum of NIRCam and MIRI without short wavelength coverage (red posterior distributions).
The input true [M/H] and log(C/O) values that were used in calculating the forward model and simulated JWST spectra are shown as blue pluses.
The incorrect high metallicity/low C/O solution can be eliminated when short wavelength data such as with NIRISS is combined with the NIRCam and MIRI data (blue posterior distributions).
While WASP-107 b does not have a bimodal posterior distributions, NIRISS improves the metallicity precision.}\label{fig:twoModeSolutions}
\end{figure*}

\section{Results}\label{sec:results}

We ran the \texttt{multinest} retrievals on the University of Arizona High Performance Computing Ocelote and El-Gato systems.
The computation of a forward model requires several seconds ($\sim$~8 seconds for an emission spectrum calculation), so \added{900 to 3,000}\deleted{many} CPU hours are required\deleted{to reach convergence: from 900 to 3,000 CPU hours} per retrieval depending on the spectrum.
We then examined the posterior distributions of derived parameters.
For the emission spectra, we include the wavelengths from 2.4~$\mu$m to 11$\mu$m for the NIRCam and MIRI instruments.
As we will show in Section \ref{sec:twoModes}, transmission spectra with NIRCam and MIRI wavelengths alone can still result in bimodal solutions for some planets so shorter wavelength data must be used in concert to obtain high precision metallicity and C/O constraints.
We therefore ran retrievals with multiple combinations of instrument modes to see which data can rule out the incorrect solution among the bimodal posterior distributions.

\subsection{Two Mode Solutions}\label{sec:twoModes}

We initially ran retrievals for transmission spectra using NIRCam and MIRI data alone (2.4 to 11 $\mu$m).
The transmission models for HAT-P-26 b and WASP-80 b can fit these longer wavelength data ($\lambda > 2.4 \mu$m) with two possible solutions: 1) a moderate metallicity, near-solar C/O ratio atmosphere, which is the true input solution in our models, or 2) a high metallicity, low C/O ratio atmosphere.
These two solutions are due to the fact that small spectral features can either be due to 1) sub-solar metallicities that have small opacities because of low abundances of absorbers or 2) high ($\gtrsim 20\times$ solar metallicities) that have small scale heights due to the high mean molecular weight of the gas \citep[e.g.][]{benneke2012retrieval,kempton2017exotransmit}.
Figure \ref{fig:twoModeSolutions} shows the posterior distributions of the metallicity and C/O parameters from the NIRCam + MIRI wavelengths.
For HAT-P-26 b and WASP-80 b, the input forward model atmospheric parameters are similar (T=990/900 K, C/O=0.55/0.55 and M/H=4.8/7.6, respectively).
In both cases, the incorrect solution arises because the high metallicity decreases the scale height, but increases the strength of the H$_2$O feature.
A very low C/O ratio $\sim$ 0.05 is needed to keep the CO absorption feature at 4.3$\mu$m from being too strong at these high metallicities.
We investigated what data was needed to ``break'' the bimodal solution into a unimodal solution by ruling out the high metallicity models.

To assess what data is needed to constrain these two mode solutions found for transmission spectra we ran 3 sets of retrievals:
\begin{enumerate}[noitemsep]
	\item NIRCam + MIRI (2.4~$\mu$m to 11~$\mu$m)
	\item HST-WFC3 + NIRCam + MIRI\label{it:HNCMI} \added{(1.1~$\mu$m to 1.7~$\mu$m and 2.4~$\mu$m to 11~$\mu$m)
}
	\item NIRISS, NIRCam and MIRI (1.0~$\mu$m to 11$\mu$m)\label{it:NINCMI}
\end{enumerate}.

For set \ref{it:HNCMI}, we use existing data from the literature described in Section \ref{sec:emisTransSpec}.
We first binned this data, where necessary, to R $\lesssim$ 40 to ensure enough points can be binned from the pre-computed correlated-K opacity grids calculated at R=100.
Either wavelength set \ref{it:HNCMI} or \ref{it:NINCMI} breaks the bimodal solution into unimodal for WASP-80 b.
However, the HST-WFC3 data added to NIRCam and MIRI (set \ref{it:HNCMI}) was not sufficient to constrain the posteriors to a unimodal solution for HAT-P-26 b, so we perform 2 additional sets of retrievals:
\begin{enumerate}[noitemsep,resume]
	\item HST-STIS + HST-WFC3 + NIRCam + MIRI\label{it:STISWNCMI}
	\item HST-WFC3 + NIRCam (higher SNR) + MIRI (higher SNR)\label{it:HighSNR}
\end{enumerate}
Set \ref{it:HighSNR} is the same as \added{Set} \ref{it:HNCMI} but the signal to noise of the NIRCam and MIRI data was artificially increased to better understand the bimodal solutions.
We set the SNR for HAT-P-26 b to be the same as for WASP-80 b (which is 1.2 K mags brighter in $K_S$) to understand if the same sets of error bars would converge on a unimodal solution.
As shown in Figure \ref{fig:twoModeSolutions}, this higher signal to noise ratio set does indeed converge to a unimodal solution.
The fact that set \ref{it:HighSNR} breaks bimodal solutions for both WASP-80 b and HAT-P-26 b suggests there is a critical signal to noise threshold, when combining with short wavelength HST data, above which the high metallicity solution is ruled out.
The HST-STIS short wavelength data does not reach this critical threshold so there is still a bimodal solution for HAT-P-26 b for case \ref{it:STISWNCMI}.

One key to breaking these bimodal solutions and correlations is to have multiple wavelength bands for the same molecule or to detect the relative strengths of the core versus wings of a spectral feature \citep{benneke2012retrieval}.
This is because the relative transit depths at different cross sections of the same absorber will scale relative to each other as a function of scale height (constraining molecular weight) but not with cloud pressure or relative abundance.

GTO observations are planned for HAT-P-26 b and WASP-107 b using NIRISS, so the final combined results will included at least NIRISS, NIRCam and MIRI data (set \ref{it:NINCMI}).
NIRISS will trace the H$_2$O bands at high signal to noise and also break the bimodal solutions.
This NIRISS + NIRCam + MIRI retrieval is shown in blue in Figure \ref{fig:twoModeSolutions}.
Once NIRISS wavelengths are added in, the metallicity is constrained to within 30\% (68\% confidence) for HAT-P-26 b and WASP-80 b.
WASP-107 b, with an input low C/O ratio, does not suffer from the bimodal problem but does benefit from the NIRISS wavelengths with improved metallicity precision by a factor of $\sim$2.
We report the final transmission retrieval results in Table \ref{tab:TranResults} for wavelength set \ref{it:NINCMI} with NIRISS, NIRCam and MIRI used together.

\begin{deluxetable*}{cccccccccc}
\tabletypesize{\scriptsize}
\tablecaption{Transmission Retrieval Results}
\tablehead{\colhead{Names} & \colhead{T} & \colhead{M/H} & \colhead{C/O} & \colhead{Amp$_{ray}$} & \colhead{Slope$_{ray}$} & \colhead{$\kappa_{Cloud}$} & \colhead{x$_{R_p}$} & \colhead{log($\gamma_1$)} & \colhead{$\kappa_{IR}$}}
\startdata
WASP-80 b & 830$^{+130}_{-230}$ & 5.6$^{+1.8}_{-1.2}$ & 0.45$^{+0.14}_{-0.16}$ & -1.7$^{+2.2}_{-2.1}$ & 3.3$^{+1.8}_{-2.1}$ & -28.89$^{+0.20}_{-0.12}$ & 0.9934$^{+0.0078}_{-0.010}$ & -0.08$^{+0.91}_{-2.8}$ & -1.72$^{+1.0}_{-0.51}$ \\
WASP-107 b & 630$^{+89}_{-130}$ & 11.8$^{+5.5}_{-3.3}$ & 0.0255$^{+0.011}_{-0.0077}$ & -1.6$^{+2.2}_{-2.1}$ & 3.2$^{+1.8}_{-1.9}$ & -28.296$^{+0.13}_{-0.074}$ & 0.908$^{+0.014}_{-0.018}$ & 0.23$^{+0.52}_{-0.61}$ & -1.36$^{+0.94}_{-0.72}$ \\
HAT-P-26 b & 885$^{+93}_{-77}$ & 4.14$^{+1.3}_{-0.97}$ & 0.51$^{+0.11}_{-0.12}$ & -2.4$^{+2.0}_{-1.7}$ & 3.1$^{+2.0}_{-2.0}$ & -29.993$^{+0.094}_{-0.11}$ & 0.855$^{+0.011}_{-0.019}$ & -0.33$^{+0.31}_{-0.55}$ & -1.35$^{+1.9}_{-0.77}$\\
\enddata
\tablecomments{Transmission spectra retrieval results using NIRISS + NIRCam + MIRI.
Error bars are the 68\% confidence intervals.}\label{tab:TranResults}
\end{deluxetable*}

\begin{deluxetable*}{cccccccc}
\tabletypesize{\scriptsize}
\tablecaption{Emission Retrieval Results}
\tablehead{\colhead{Names} & \colhead{log($\gamma_1$)} & \colhead{log($\gamma_2$)} & \colhead{$\kappa_{IR}$} & \colhead{$\alpha$} & \colhead{T} & \colhead{M/H} & \colhead{C/O}}
\startdata
HD 189733 b & -1.21$^{+0.62}_{-0.47}$ & -1.21$^{+0.62}_{-0.48}$ & -1.551$^{+0.089}_{-0.055}$ & 0.49$^{+0.32}_{-0.32}$ & 1176$^{+17}_{-24}$ & 5.79$^{+1.3}_{-0.98}$ & 0.687$^{+0.041}_{-0.044}$ \\
WASP-80 b & -1.28$^{+1.1}_{-0.48}$ & -1.29$^{+1.1}_{-0.48}$ & -1.483$^{+0.21}_{-0.079}$ & 0.50$^{+0.37}_{-0.37}$ & 903$^{+13}_{-37}$ & 8.9$^{+3.5}_{-2.3}$ & 0.672$^{+0.066}_{-0.069}$ \\
HAT-P-19 b & -1.18$^{+1.2}_{-0.53}$ & -1.15$^{+1.2}_{-0.54}$ & -1.38$^{+0.39}_{-0.23}$ & 0.52$^{+0.33}_{-0.35}$ & 980$^{+43}_{-71}$ & 9.6$^{+16}_{-5.0}$ & 0.40$^{+0.23}_{-0.19}$ \\
GJ 436 b & -1.32$^{+0.72}_{-0.45}$ & -1.31$^{+0.70}_{-0.46}$ & -1.316$^{+0.14}_{-0.10}$ & 0.50$^{+0.34}_{-0.33}$ & 700$^{+13}_{-19}$ & 1000$^{+300}_{-190}$ & 0.243$^{+0.052}_{-0.064}$\\
\enddata
\tablecomments{Emission retrieval results using NIRCam + MIRI.
Error bars are the 68\% confidence intervals.}\label{tab:EmisResults}
\end{deluxetable*}

\subsection{Retrieved Mass-Metallicity relation}\label{sec:retrievedMassM}

\begin{figure*}
\begin{centering}
\includegraphics[width=0.75\textwidth]{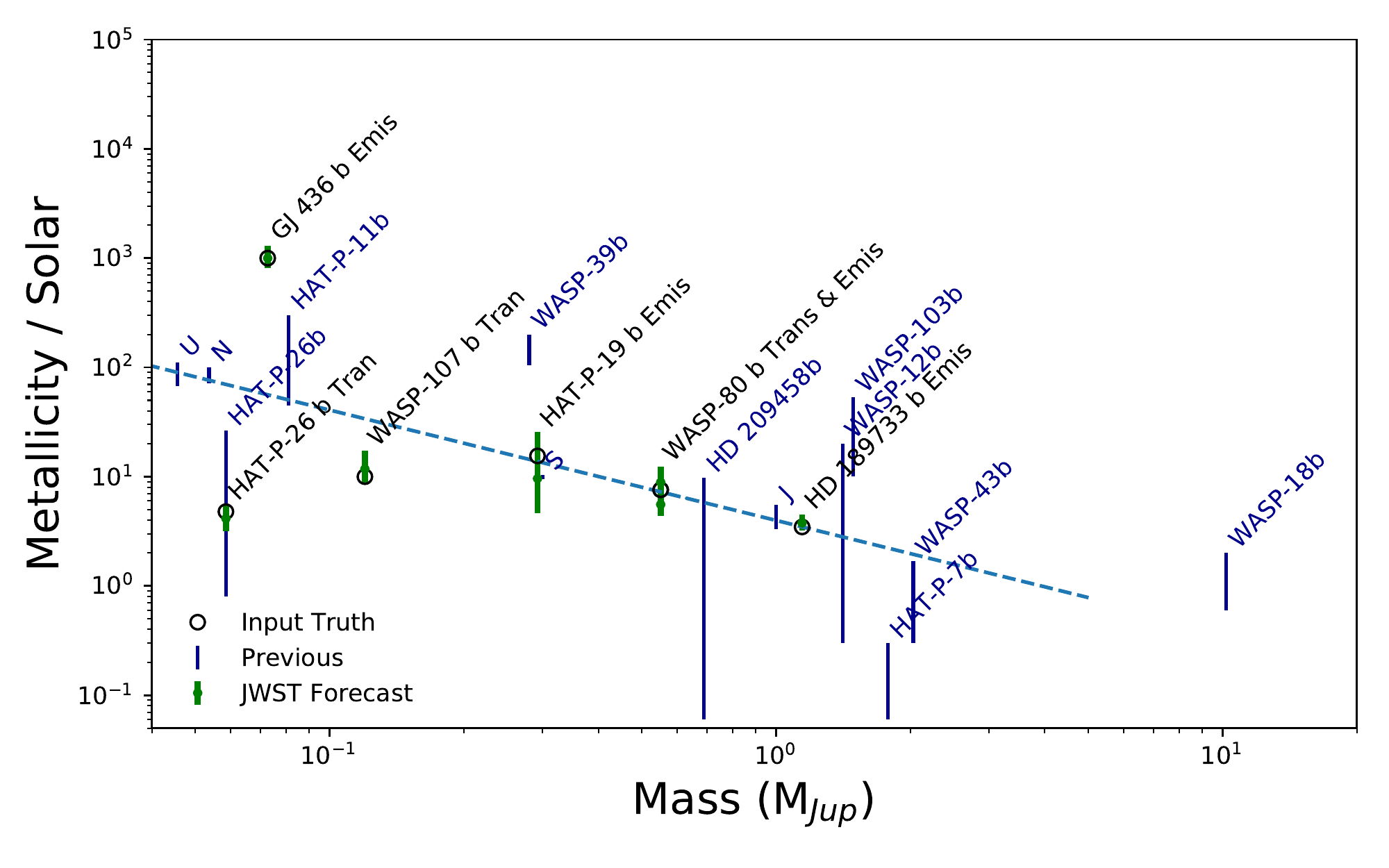}
\caption{Input metallicities (open circles) for the forward models were assigned to be consistent with literature spectra for the masses $<0.2 M_{Jup}$ or the mass-metallicity relation from \citet{kreidberg2014wasp43} (dashed line) for the masses $> 0.2 M_{Jup}$.
These initial spectra were assigned Gaussian noise from JWST sensitivities and then the metallicity was retrieved with the CHIMERA model suite.
The JWST errors for transmission and emission spectra are sufficient to measure deviations from a power law mass-metallicity relation to high statistical significance.
The transmission model metallicity precisions plotted here include NIRISS observations to help constrain bimodal solutions discussed in Section \ref{sec:twoModes}.
We include solar system observations as well as constraints from the literature using Spitzer and HST, discussed in Section \ref{sec:retrievedMassM}.
\added{We note that these results combine metallicities derived from CH$_4$, H$_2$O and self-consistent modeling, while assuming Solar ratios of elements heavier than helium (except for partitioning oxygen and carbon by the C/O ratio) and chemical equilibrium.}
}\label{fig:massMetal}
\end{centering}
\end{figure*}

Figure \ref{fig:massMetal} shows the input mass-metallicity relation discussed in Section \ref{sec:abundMetal}.
The general trend of decreasing metallicity with increasing mass in observations \citep{kreidberg2014wasp43} can be explained by the core accretion paradigm.
At increasing planetary H/He envelope mass, any planetesimals accreted during the planet-formation era have an ever larger mass of total H/He for their delivered metals to be diluted within.
JWST will allow broad studies of many planets at high metallicity precision to understand if this general trend in Solar System planets also holds true for exoplanets.
Figure \ref{fig:massMetal} also shows the metallicities in the literature compiled by \replaced{\citet{kreidberg2014wasp43}, as well as}{\citet{mansfield2018hatp7}, which includes} individual metallicity measurements from \citet{fraine2014hatp11}, \citet{wakeford2017hatp26}, \citet{kreidberg2015wasp12}, \citet{line2016hd209wfc3}, \citet{arcangeli2018wasp18}, \citet{wakeford2018wasp39}, \added{\citet{kreidberg2014wasp43} and \citet{kreidberg2018wasp103}} and mass measurements from \citet{bakos2010hatp11}, \citet{hebb09}, \citet{triaud2010spinOrbit}, \citet{faedi2018wasp39}, \citet{gillon2012wasp43}, \added{\citet{pal2008hatp-7} and \citet{gillon2014wasp103}}.
The WASP-39 b atmospheric metallicity has especially high precision because it uses \replaced{constraints with self-consistent modeling of the energy budget and chemistry}{a continuous wavelength coverage from 0.3~$\mu$m to 1.7~$\mu$m as well as {Spitzer} IRAC photometry.}

\added{For these measurement in the literature, we have combined together results that come from CH$_4$ constraints (Solar System), self-consistent modeling (WASP-18 b, HAT-P-7 b, WASP-103 b) and H$_2$O constraints (remaining measurements).
In order to combine these three measurements of elemental abundances in a single plot, we assume that all the elemental abundances that are heavier than Helium are at fixed ratios to each other at the same values as the Sun.
Thus, the elemental abundances can be parameterized by (X/H)/(X/H)$_S$ where X is the abundance of an element, H is the abundance of Hydrogen and (X/H)$_S$ is the solar abundance ratio per Hydrogen atom.
Then, the total carbon and oxygen content is partitioned into carbon and oxygen based on the C/O ratio.
If the planet atmosphere's C/O is unknown, then a solar value of C/O=0.55 is assumed or marginalized over.
From these elemental abundances, it is assumed that the gas is in chemical equilibrium and thus partitions into the molecules of H$_2$O, CH$_4$, CO, CO$_2$, NH$_3$ etc.
For the majority of exoplanets measurements in the literature and the forecasted JWST measurements in this work, the atmospheres are modeled with an overall metallicity parameter and the confidence intervals are returned by marginalizing over all other variables.
}

The forward models are assigned to follow the \citet{kreidberg2014wasp43} relation with the exception of HAT-P-26 b, GJ 436 b and WASP-107 b, which have metallicities tuned to match literature spectra of the objects since they already show deviations from the Solar System trend.
The metallicity of each forward model is plotted as an open circle in Figure \ref{fig:massMetal}.
The 68\% confidence levels for the retrieved metallicities are shown as error bars for each planet model \added{with values available in Tables \ref{tab:TranResults} and \ref{tab:EmisResults}}.
These confidence intervals can nearly approach solar system precisions under our model assumptions (which include low cloud opacities) when NIRISS, NIRCam and MIRI are used simultaneously in transmission or NIRCam and MIRI are used simultaneously in emission.
It should be noted that population synthesis models can produce a spread in metallicities for a \replaced{give}{given} mass, especially for Neptunes and Super-Earths \citep{fortney2013lowMassP}, so there is value in measuring the deviations from Equation \ref{eq:MassMetal} to better understand planet formation conditions and evolution.

\begin{figure}
\begin{centering}
\includegraphics[width=1.08\columnwidth]{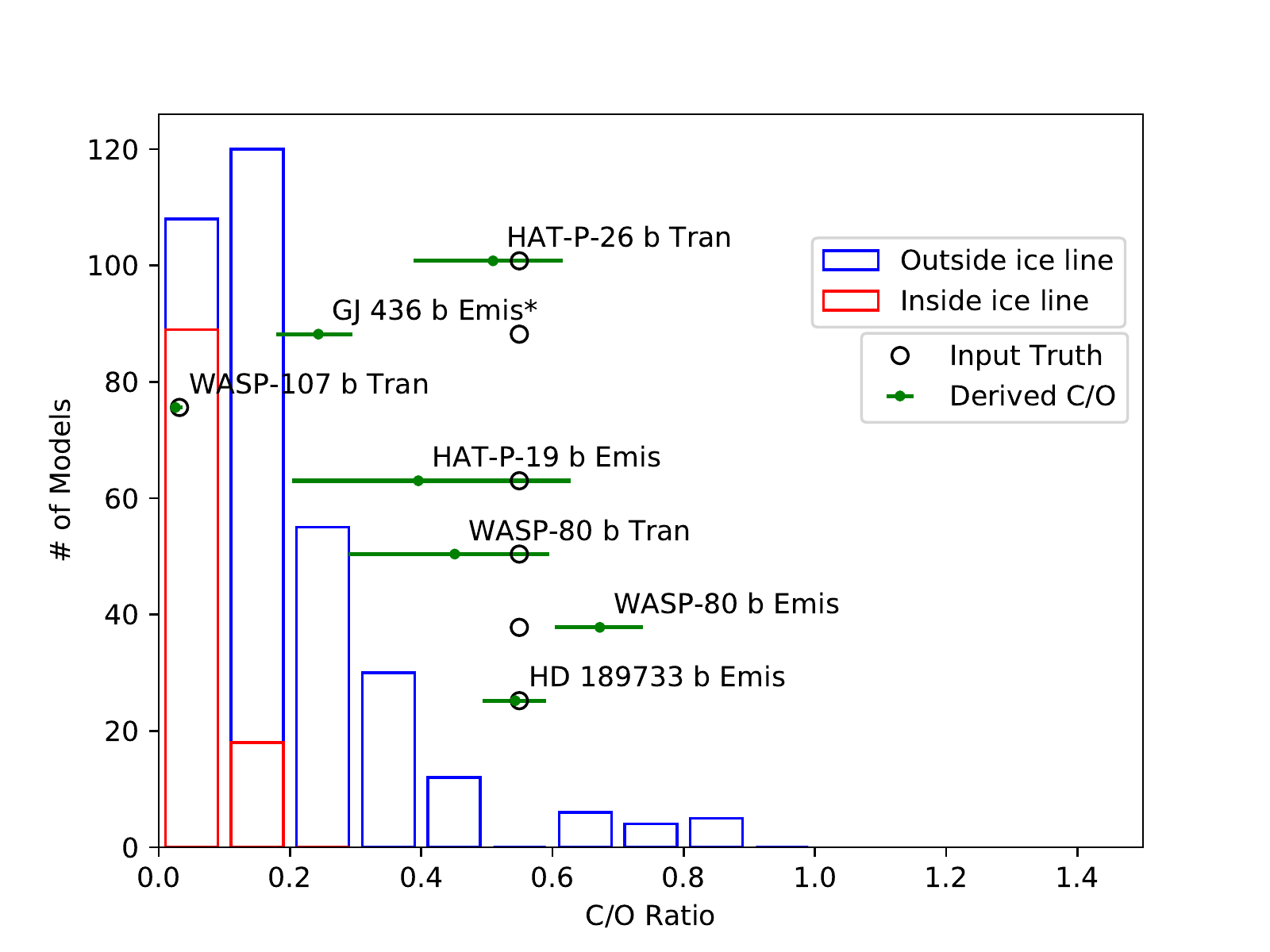}
\caption{Input C/O (open circles) for the forward models were assigned to be consistent with literature spectra (WASP-107 b) or otherwise set to the Solar C/O=0.55.
The retrieved 68\% confidence intervals (green error bars) are shown for JWST observations with NIRISS, NIRCam and MIRI using the CHIMERA model suite \added{at arbitrary Y positions}.
Example sets of chemical models (ones that have carbon-deficient inner disks) from \citep{mordasini2016planetFormationSpec} are shown as histograms over a range of disk abundance ratios for volatiles and refractories.
The retrieved confidence intervals are small enough to test planet formation predictions (under specific disk assumptions) such as these where planetesimal accretion drives all C/O ratios below unity.
Note that these histograms do not represent population syntheses but rather a variety of different assumed chemistries.
*The retrieved C/O for GJ 436 b does not match the input because chemical quenching is only included in the forward model, so the spectrum is fit with a low C/O ratio.
}\label{fig:CtoORatios}
\end{centering}
\end{figure}

\subsection{Retrieved C/O Ratios}\label{sec:retrievedCtoO}

C/O ratios in planetary atmospheres can potentially be used to characterize how a planet forms or where within a disk it forms \citep{oberg11}.
We show the retrieved C/O ratios for our \replaced{Multinest}{\texttt{multinest}} retrievals in Figure \ref{fig:CtoORatios}.
Our initial forward models all have a C/O of 0.55 except for WASP-107 b as listed in Table \ref{tab:gtoSources}.
WASP-107b requires an unusually low C/O ratio to suppress the CH$_4$ features, as measured by the HST WFC3 spectrum.
Retrievals can constrain C/O ratios to better than 50\% accuracy.
This is well below the spread in different possible chemistries of disks.

Figure \ref{fig:CtoORatios} shows a subset of the chemical models considered in \citet{mordasini2016planetFormationSpec}.
The planet formation models shown in Figure \ref{fig:CtoORatios} are for disks that can have a range of carbon to silicate ratios, refractory to volatile ratios and the presence or absence of clathrates but always include a depletion of carbon towards the central star.
This assumption of carbon depletion, motivated by observations of abundances in the Solar System, causes planets to form and evolve to have atmospheric C/O ratios all less than 0.9 even for large ranges of carbon, silicon, volatiles and clathrates considered.
This is because accreting planetesimals (made of ices or refractories that are high in oxygen and low in carbon) will become incorporated in the atmosphere and lower the C/O ratios.
If the inner disk is not depleted in carbon, it is possible to produce planets with C/O ratios large than 1, but this would require a different kind of disk than observed in the Solar System.
If such disks are common, it may be possible to determine the formation location of a planet from its atmospheric C/O ratio.
We will be able to test the prediction from these disk models to high precision with JWST NIRCam and MIRI observations and retrievals shown in Figure \ref{fig:CtoORatios}.

GJ 436 b's spectral retrieval does not match the input solar C/O ratio of 0.55.
This is because our input forward model has chemical quenching at 0.32 bar shown in Table \ref{tab:gtoSources} as found in \citet{morley2017gj436bRetrieval}.
Our spectral retrieval assumes a fixed chemical quench point of 10$^{-4}$ bar, so it essentially has no chemical quenching.
We fixed this parameter in the retrieval for two reasons: 1) to reduce the number of free parameters and reduce computation time and 2) to determine the degree to which non-equilibrium chemistry can be detected in an atmosphere by assessing how well this model performs.
The lack of quenching forces the model to an abnormally low C/O ratio of 0.24$^{+0.05}_{-0.06}$ which also suppresses the CH$_4$ feature at 3.6~$\mu$m.

\begin{figure*}
\begin{centering}
\includegraphics[width=0.8\textwidth]{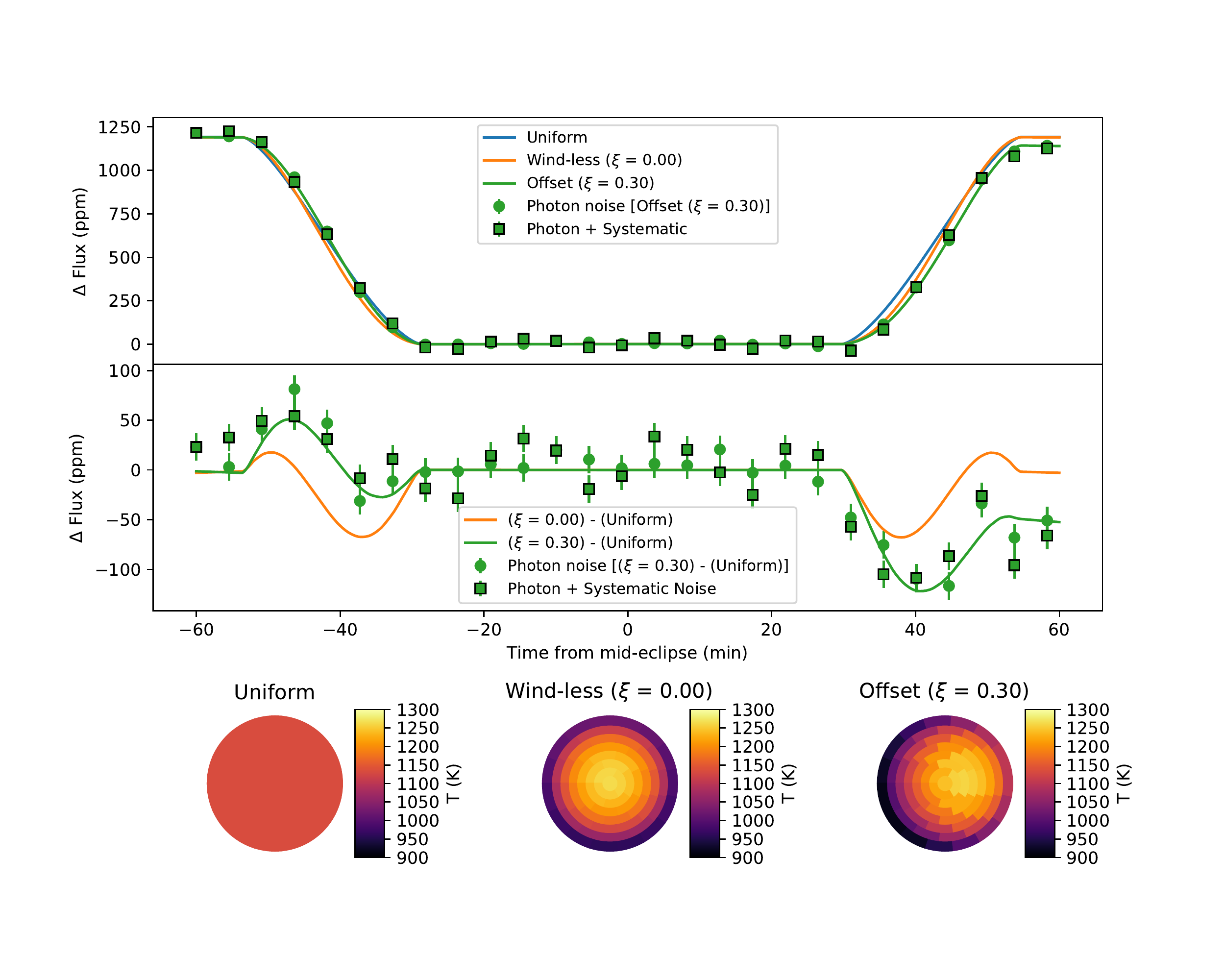}
\caption{Secondary eclipses for 3 different surface maps using \texttt{spiderman} \citep{louden2017spiderman} for the F444W filter.
We compare a uniform surface, centrally located spot and hotspot offset using the prescription from \citet{zhang2017bulkCompDynamics}.
The top panel shows the change in flux relative to the star, the middle panel shows the differential flux between the non-uniform models and the uniform model and the bottom panel shows the brightness maps of the three models.
The signals are up to 67 ppm during the secondary ingress and 122 ppm during secondary egress.
The simulated data are shown for the case with no systematic and with a sinusoidal systematic error applied to the hotspot offset case.}\label{fig:eclipseMap}
\end{centering}
\end{figure*}

\section{Eclipse Mapping}\label{sec:EclipseMap}

Bright targets like HD 189733 b will have high signal to noise ratios per integration (SNR=7000 per 2.4 second integration when binning over all F444W wavelengths) permitting eclipse mapping as the planet's disc is occulted by the star's limb during ingress and egress \citep{rauscher2009eclipseMapping}.
\citet{majeau2012eclipsemap189} and \citet{deWit2012eclipsemap189} used the Spitzer IRAC 8~$\mu$m light curve to create a map of the surface.
As predicted by atmospheric circulation models \citep[e.g.][]{showman2002circulation51peg}, HD 189733 b's 8~$\mu$m hot spot is shifted eastward from the sub-stellar point \citep{knutson2007map189}.
The ingress and egress light curves from the Spitzer IRAC 8~$\mu$m deviate from a uniform disk by 6~$\sigma$.
Furthermore, \citet{majeau2012eclipsemap189} and \citet{deWit2012eclipsemap189} put the first constraints on the equator-to-pole temperature contrasts.
The equator-to-pole differences are not measurable from phase curves of transiting tidally locked planets \citep[e.g.][]{knutson2012phase189} since the rotational axis is nearly perpendicular to the line of sight.

We explore the level of eclipse mapping possible using one eclipse observation with JWST's NIRCam F444W mode.
We use a simple prescription from the appendix of \citet{zhang2017bulkCompDynamics} to describe the 2D spatial map of the planet's surface and \texttt{spiderman} \citep{louden2017spiderman} to calculate the eclipse curve.
The map model is a function of the orbital parameters, the ratio of the radiative timescale to advective timescale ($\xi$), the dayside/nightside temperatures and the stellar temperature.
We consider 3 example cases in Figure \ref{fig:eclipseMap}:
\begin{enumerate}[noitemsep]
	\item A uniform map with no night/day temperature contrast
	\item A wind-less map with $\xi = 0$ that has a spot centered at the sub-stellar point
	\item A wind model with $\xi = 0.3$ that has an eastward hot spot offset.\label{it:advectiveModel}
\end{enumerate}
Case \ref{it:advectiveModel} represents a value similar to the $20\deg$ phase offset observed for the 4.5$\mu$m Spitzer phase curve of HD 189733 b \citep{knutson2012phase189}.

The three models' eclipse light curves are nearly indistinguishable to the eye in Figure \ref{fig:eclipseMap} so we also show the differential flux between these models.
The signal of non-uniform brightness can be seen with an S-shaped differential light curve with an amplitude of up to 68ppm.
Further, the hot spot offset introduces a discernible asymmetry between the ingress and egress, with a maximum signature of 122 ppm.
The phase curve of the offset model introduces a slope to the baseline (out-of-transit light curve) which will likely be removed in the data analysis, as there can be trends in the light curve due to stellar activity or systematic noise effects.
Full phase curve observations of the whole orbit will also provide better constraints of the planet and stellar surface inhomogeneities.

We calculate the photometric error (including background and read noise) for the broadband integrated flux in the F444W band.
In Figure \ref{fig:eclipseMap}, we show the resulting error that can be obtained in this photon limit, which is 14 ppm for 4 minute bins (100 integrations each).
In reality, there may be a systematic noise floor of around 30 ppm that prevents measurements below this level.
To simulate the effect of a systematic noise floor, we introduce a sinusoidally varying signal that has an amplitude of 30 ppm and a period of 15 minutes to the error and show this case as well.
Even with the pessimistic 30 ppm systematic noise, a deviation from a non-uniform surface for both a wind-less case 2 and hotspot offset case 3 are readily detectable.

We created a simulated time series with $\xi=0.3$ and a day/night temperature contrast of 500 K  with random Gaussian photon noise and a 30 ppm sinusoidal systematic (green data points with square symbols in Figure \ref{fig:eclipseMap}).
This time series was fit with an MCMC sampler \citep{foreman-mackey2013emcee}.
The radiative timescale parameter $\xi$ can be constrained to $\pm$ 0.07, which corresponds to an angular precision of the hotspot offset of $\pm \sim 3.5\degree$ of longitude.

\section{Conclusion}\label{sec:Conclusion}
We have outlined the observations and models of a NIRCam + MIRI GTO survey of planets cooler than $\lesssim$1150~K, which are in the temperature regime where methane (CH$_4$) can become a significant absorber.
These include  HD~189733~b, WASP-80 b, HAT-P-19 b, WASP-107 b, GJ 436 b and HAT-P-26 b.
When combining results with the NIRCam, MIRI and NIRISS instruments to cover the $\sim$1 to 11$\mu$m wavelength range, this survey permits precision measurements of planet metallicity ($\pm$20\% to 170\%) and the C/O ratio ($\pm \sim$10\% to 60\%) in these planet atmospheres.
These precisions from random uncertainties are unprecedented.
However, we may be limited by larger systematic effects, exemplified by the illustrative case of carbon quenching in GJ 436 where the retrieved model fails to recover the input C/O ratio from the forward model if it does not include carbon quenching.
This survey covers planet masses from 20 to 360~M$_\oplus$ and enables a high precision study of the mass-metallicity relation, which has only preliminary trends gleaned from Solar System and exoplanet measurements.

We find that using NIRCam and MIRI alone to measure the transmission spectrum of a moderate-metallicity, near-solar C/O planet can result in bimodal solutions.
These bimodal solutions found for HAT-P-26 b and WASP-80 b are either a high-metallicity--low C/O ratio solution (M/H $\approx 10^2$, C/O $\approx 10^{-2}$) or a moderate metallicity--moderate C/O ratio solution (M/H $\approx 1$,  C/O $\approx$ 0.6).
In the case of WASP-80 b, HST WFC3 data can break these bimodal solutions but HAT-P-26 b requires higher signal to noise data (such as with the JWST NIRISS instrument) to pick the correct unimodal solution.
\added{This is similar to retrieval results for GJ 1214~b, which show that existing optical and near-infrared data can be fit with a wide range of cloud-top pressures; the cloud top uncertainty adversely affects the derived H$_2$O abundance but these solutions will be better-constrained with a MIRI LRS spectrum \citet{barstow2015jwstSystematics}.}

The retrieved atmospheric C/O ratios can potentially be used to understand planet formation\deleted{models}.
If planet formation processes preserve the history of their formation such as in the models from \citet{mordasini2016planetFormationSpec} and \citet{oberg11}, it is possible to test specific model predictions.
In sets of models with disk chemistry consistent with Solar System measurements, solid planetesimals have low C/O ratios because they are either made of ice or of refractory elements where carbon has been vaporized.
Planetesimal impacts with a forming planet will thus lower the C/O ratios observed in present-day atmospheres \citep{mordasini2016planetFormationSpec}.
JWST's NIRCam and MIRI instruments provide the high precision abundance measurements of carbon-bearing molecules to test these C/O hypotheses.

We simulated the precisions achievable with eclipse mapping on our brightest system HD 189733 b.
A single eclipse is sufficient to measure the hotspot offset, as parameterized with an advective to radiative timescale parameter $\xi$ in \citet{zhang2017bulkCompDynamics}.
Even in the presence of correlated sinusoidal noise at the $\sim$ 30 ppm level, the hotspot is detectable and can be constrained to within a few degrees of longitude.

\acknowledgments
\section*{acknowledgements}
We thank our collaborators J. Bean, J. Lunine, P.-O. Lagage, J. Bouwman for helpful discussions on observational modes as well as coordinations of GTO programs.
Thanks for J. Bean's very helpful comments that improved the manuscript and for M. Mansfield for providing atmospheric metallicities from the literature.
The funding for NIRCam team members, including E. Schlawin and M. Rieke, is provided by NASA Goddard.
TP Greene acknowledges funding support from the NASA JWST program WBS 411672.
\added{We thank the anonymous referee for the valuable suggestions and references to improve this work.}

%

\vspace{5mm}
\facilities{JWST(NIRCam), JWST(MIRI)}


\software{\texttt{astropy} \citep{astropy2013}, 
          \texttt{multinest} \citep{feroz2009multinest},
          \texttt{batman} \citep{kreidberg2015batman},
          \texttt{spiderman} \citep{louden2017spiderman},
          \texttt{pynrc} \url{https://github.com/JarronL/pynrc},
          \texttt{emcee} \citep{foreman-mackey2013emcee}
          }

\bibliographystyle{apj}
\bibliography{this_biblio}



\end{document}